%% file: main.tex
\documentclass[11pt,onecolumn]{article}

\usepackage[margin=1in]{geometry}
\usepackage{amsmath,amssymb}
\usepackage{graphicx}
\usepackage[hidelinks]{hyperref}
\usepackage{xcolor}
\usepackage{lineno}
\usepackage{booktabs}
\usepackage{siunitx}
\usepackage{caption}
\usepackage{placeins}
\newcommand{\orcidicon}[1]{\href{https://orcid.org/#1}{\textcolor[HTML]{A6CE39}{\,\textbullet\,}}}

\title{Schedule-dependent basin occupation and instance-specific effective scales in cycled reverse annealing on programmable quantum annealers}

\author{Luis Lozano\,\orcidicon{0000-0001-7202-3437}\\ \small Tecnol\'ogico de Monterrey, Campus Santa Fe \\ \small \texttt{lalozanom@tec.mx}}
\date{\today}

\begin{document}
\maketitle

\begin{abstract}
Reverse annealing initializes a programmable quantum annealer in a chosen classical state, reintroduces quantum fluctuations down to a pause point, and returns to readout, and repeated cycles of this protocol constitute a configurable dissipative dynamics whose control knobs --- pause depth, pause duration, schedule shape --- are routinely treated as interchangeable with a temperature. On mixed-frustration 12-qubit Ising instances run on two D-Wave generations (Advantage2, Zephyr topology, and Advantage\_system6.4, Pegasus topology), we measure how schedule shape redistributes the subsystem readout distribution, and we calibrate the effective scales that each schedule realizes, using exhaustive enumeration of the $2^{12}$ configurations, a locked analysis plan archived with its prior-knowledge disclosure, and classical reference families regenerated with per-chain random streams. Three main results are established, followed by two corrective negative findings. First, schedule shape redistributes basin occupation: under a block-permutation test with false-discovery control, 11 of 20 random instances respond to a change from a single-pause to a two-pause schedule (15 of 20 for a long-hold schedule), with dominant-configuration shifts up to 38 percentage points and, on several instances, a change in the identity of the dominant configuration. Second, the single-qubit probe temperature does not transfer to the many-body problem: the instances' moment-matched inverse temperatures span $0.63$--$9.1$ against a probe value of $7.22$, and pause depth and pause duration each shift the moment-matched scale on most instances (19/20 and 15/20 respectively, in a factorial decomposition). Third, no processor observation in this dataset requires more than the tested classical families: with the update-budget axis of the classical references extended below one sweep per cycle, every one of 33 (device, instance, schedule) cells is reproduced within the tested classical family (minimum intersection--union $p = 0.76$), and a spin-vector Monte Carlo model running the actual schedules brings the late-window occupation statistics within semiclassical reach as well. A pre-registered linear predictor of schedule sensitivity fails on held-out instances under both the original and a corrected conditioning, and an earlier pilot's report that a two-pause schedule doubles a quantum-over-classical inversion rate is contradicted: under corrected conditioning the two-pause rate is significantly below the pilot value (Fisher $p = 0.0013$). Reverse-anneal schedules act as instance-specific basin-occupation controls whose effective energy and update scales must be calibrated per instance and per schedule; none of the observations reported here licenses a quantum-mechanistic reading.
\end{abstract}

\section*{Introduction}

Programmable quantum annealers~\cite{albash2018adiabatic,crosson2021prospects,kendon2026qacm} are routinely used as samplers for statistical-physics problems~\cite{king2022coherent,king2023critical,king2024false_vacuum,teza2025finite,aronoff2026coherence}. Their forward-anneal output admits a reasonably understood interpretation as biased Boltzmann-like sampling from the programmed Ising Hamiltonian at a device-specific effective temperature~\cite{vuffray2022boltzmann,grattan2025thermometry}, and the open-system dynamics behind that output have been characterized within a Schr\"odinger--Lindblad--Markovian hierarchy~\cite{mehta2025dwave}. The reverse-anneal protocol --- initialize in a prepared classical state at $s=1$, ramp to a pause point $s_p<1$ where quantum fluctuations are reintroduced, hold for a pause time $t_p$, return to $s=1$ for readout --- has been proposed both as a local-exploration heuristic and as a probe of relaxation~\cite{dwave_reverse_anneal_docs,marshall2019pausing,chen2020pausing,mehta2025unraveling,pelofske2025erasing}, and recent work uses its turning point as a thermodynamic control parameter~\cite{pawlowski2026thermal} or maps its parameter regimes for optimization~\cite{menger2026reverse}.

Two questions motivate this study. First, does the \emph{shape} of the reverse-anneal schedule control the final sampling distribution beyond a rescaling of effective temperature? Second, at what effective scales --- energy scale and update budget --- do classical reference processes reproduce the cycled dynamics that the device realizes? The answers matter for benchmarking, for the interpretation of QPU samples as thermal-like, and for any application that treats the schedule as a tunable knob.

A companion study of subsystem relaxation on the same working point~\cite{locth1_companion} reported, from a 40-instance pilot, an interpretation in which a two-pause schedule ($0.45 \to 0.35$) approximately doubled the fraction of instances whose subsystem readout preferred a quantum reduced-Gibbs target over a matched classical target. The present paper tests that interpretation on a fourfold larger dataset with a cycled-protocol observable, exhaustively enumerable instances, and classical reference families scanned over their nuisance parameters.

The paper makes four contributions. First, a gauge-correct atlas of schedule response on twenty random mixed-frustration instances: 11 of 20 respond to a single-pause $\to$ two-pause change under a block-permutation test with Benjamini--Hochberg control~\cite{benjamini1995fdr}, with dominant-weight shifts up to 38 percentage points and identity changes of the dominant configuration on two instances (five under long-hold). Second, an effective-scale calibration of the protocol: per (instance, schedule) moment-matched inverse temperatures $\beta_\mathrm{mm}$, distribution-fit scales $\hat\beta$, and empirical update budgets $n_\mathrm{eff}$, with a factorial decomposition showing that pause depth and pause duration each move $\beta_\mathrm{mm}$ on most instances. Third, a classical-family null: against classical reference families whose update budget is scanned continuously --- including below one sweep per cycle, a regime coarse budget grids omit --- every cell in the dataset is reproduced within the tested family (a broad family, so this is a bounded cellwise nonrejection, not a model-class sufficiency claim), and schedule-aware spin-vector Monte Carlo~\cite{shin2014svmc} covers the late-window statistics that sweep-based samplers miss. Fourth, two negative results: the pre-registered structural predictor fails held-out validation under both tested conditionings, and the companion pilot's two-pause enhancement is contradicted with $p = 0.0013$.

\section*{Protocol}

\paragraph{Subsystem--environment partition.}
We work on a D-Wave Advantage2 Zephyr-topology QPU (solver \texttt{Advantage2\_system1}) with cross-generation comparisons on Advantage\_system6.4 (Pegasus): random 3-regular logical graphs with $N=12$ qubits, a 4-qubit connected subsystem $S$, an 8-qubit environment $E$, boundary-coupling strength $\lambda=0.5$, internal couplings ferromagnetic with bond-frustration probability $p_S=0.5$ applied only to the $S$--$S$ edges, and random longitudinal fields $h_i \sim \mathcal{U}[-W, W]$ with $W=1.0$. These parameters reproduce, unchanged, the mixed-frustration working point of the companion pilot study~\cite{locth1_companion}, so that the present, fourfold larger dataset is directly comparable to it. The environment is \emph{initialized} only: no field pins $E$ during the anneal, and on several instances $E$ leaves its initial configuration in most cycles (Sec.~\nameref{sec:conditioning}). All instances are archived as explicit $(h, J, S, E)$ dictionaries with exhaustive-enumeration fingerprints; seeds are not used as instance definitions.

\begin{figure}[!htbp]
\centering
\includegraphics[width=0.98\textwidth]{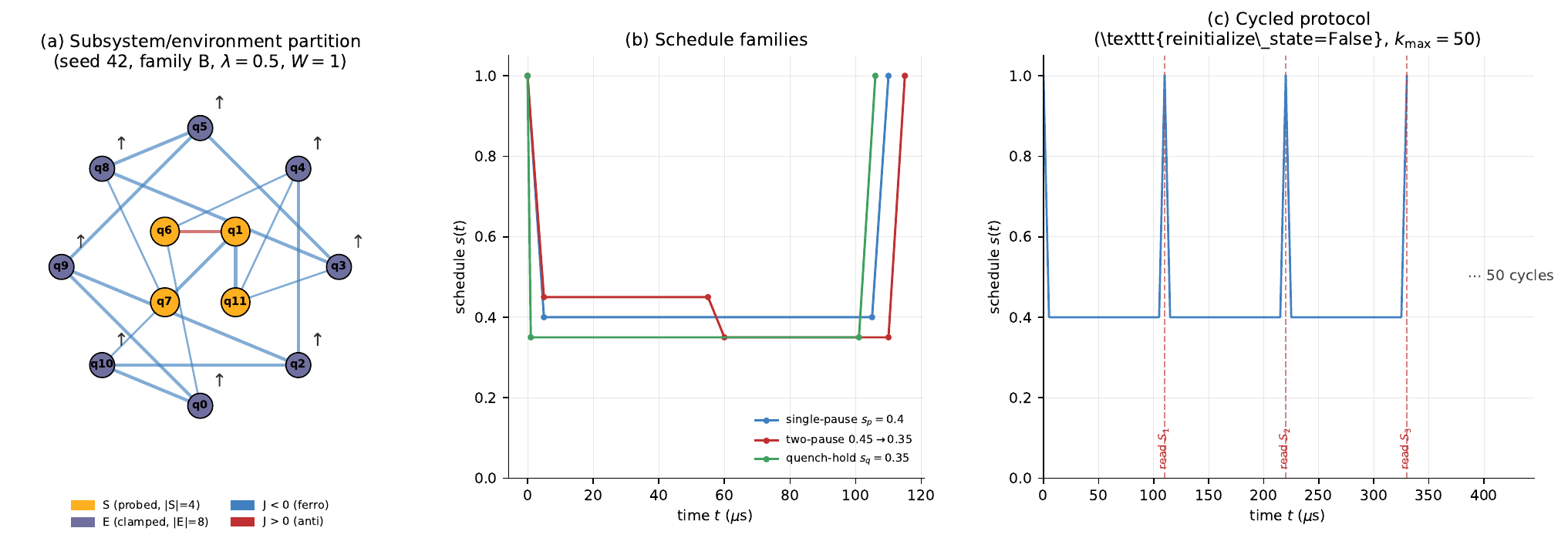}
\caption{\textbf{Experimental protocol.}
(a) Subsystem/environment partition on a representative instance
(seed 42, family B; $\lambda=0.5$, $W=1$): four subsystem qubits $S$ (orange)
are probed while eight environment qubits $E$ (grey) are \emph{initialized} to
a product configuration (arrows); no additional field constrains $E$ during
the anneal. Edges color the longitudinal couplings $J_{ij}$ (blue $J<0$, red
$J>0$), with line width proportional to $|J_{ij}|$.
(b) Schedule-family comparison: single-pause reverse anneal at $s_p=0.4$,
two-pause reverse anneal $s_1=0.45 \to s_2=0.35$, and quench-and-hold at
$s_q=0.35$. (c) Cycled reverse-anneal protocol: a single submission of
$K=50$ anneals with \texttt{reinitialize\_state=False} produces a
$K$-step chain through the reverse-anneal operator; the subsystem
configuration is read out after every cycle (red ticks).}
\label{fig:fig1}
\end{figure}

\paragraph{Schedule families.}
Four schedule families were acquired (Fig.~\ref{fig:fig1}): single-pause reverse anneals over $s_p \in \{0.35, 0.40\}$ and $t_p \in \{100, 500\}\,\mu\mathrm{s}$ (a full $2\times 2$ factorial), a two-pause reverse anneal with $(s_1, s_2) = (0.45, 0.35)$ and $50\,\mu\mathrm{s}$ at each depth, a quench-and-hold control, and a forward-anneal control. The cycled acquisitions use three of these: the baseline single-pause ($s_p=0.4$, $100\,\mu$s), the long-hold single-pause ($s_p=0.35$, $500\,\mu$s), and the two-pause schedule. Ramp times are $\tau_\mathrm{in}=\tau_\mathrm{out}=5\,\mu$s throughout, so the readout state is the state at the freeze-out crossing during the common ramp-out rather than the state at the pause point; differential comparisons between schedules are unaffected, but ``pause point'' labels the schedule, not the measured ensemble. In the cycled dataset, pause depth and pause duration co-vary between the three schedules; the endpoint dataset's $2\times2$ factorial is what licenses their separation (Sec.~\nameref{sec:scales}).

\paragraph{Cycled reverse-anneal protocol.}
Setting \texttt{reinitialize\_state=False}~\cite{dwave_reverse_anneal_docs} makes each of \texttt{num\_reads} consecutive anneals start from the previous anneal's readout, so a single submission of $K=50$ reads produces a 50-step chain through the reverse-anneal operator: the state latched at readout seeds the next anneal. Jobs were submitted with \texttt{answer\_mode="raw"} (histogram mode would destroy cycle ordering). One programming cycle covers each 50-cycle chain (measured \texttt{qpu\_programming\_time} $= 34.6$\,ms per job), so all cycles within a chain share one realization of the programmed Hamiltonian $H+\delta H$, where $\delta H$ is the intrinsic control error; consecutive cycles are separated by readout (median $41.4\,\mu$s in this acquisition) and $19.1\,\mu$s of delay at $s=1$; the solver timing block and the schedule waypoints of every condition are recorded in the per-acquisition metadata sidecars of the release. For each (instance, schedule) we ran 30 chains from each of three initial states ($S_\uparrow$, $S_\downarrow$, $S_\mathrm{random}$; $E$ all-up), i.e.\ 90 chains per cell.

\paragraph{Readout noise floor.}
Because a readout error under \texttt{reinitialize\_state=False} propagates into the next cycle, the frozen cells of the dataset provide a device-level bound: across the seven fully frozen cells (630 chains), zero chains showed any subsystem transition over 49 cycles, giving a Clopper--Pearson 95\% upper bound of $0.58\%$ on the per-chain spurious-transition probability. The mobile-cell signals analyzed below exceed this floor by more than an order of magnitude and are not attributable to readout error.

\paragraph{Effective temperature: a single-qubit calibration only.}
Single-qubit probes with bias $h=0.5$ under reverse annealing~\cite{grattan2025thermometry} give $\beta_\mathrm{probe}$ between $6.6$ and $7.8$ across probes and pause depths, with a per-probe counting uncertainty $\delta\beta \approx 0.27$--$0.50$ dominated by minority counts of 4--14 in $10^4$ reads; a naive $\pm 0.06$ (the spread of three point estimates, two of which returned byte-identical counts) understates this. The probes ran with \texttt{reinitialize\_state=True}, unlike the experiment, and on qubits disjoint from every experiment qubit. We therefore treat $\beta_\mathrm{probe}$ as an operational single-qubit calibration and never as a many-body matched temperature; the many-body scales are carried by the per-instance quantities $\beta_\mathrm{mm}$ and $\hat\beta$ defined below. This distinction matters quantitatively: at $\beta = 7.22$ the exact classical equilibrium subsystem collision probability rounds to $1.0000$ on 8 of the 20 training instances --- including flagship instances 8192 and 14029 --- and exceeds $0.95$ on 14 of 20, so on such instances a classical reference at the probe temperature is guaranteed to disagree with any mobile observation regardless of the physics.

\paragraph{Observables.}
We use the subsystem readout distribution $P_S$; the memory order parameter $M(k)$ (maximum pairwise total-variation distance between subsystem marginals from different initial states at cycle $k$); the preparation-survival probability $S(49) = \Pr[\sigma_S(49)=\sigma_S(0)]$, a per-chain Bernoulli statistic (with $K=50$ this quantity admits exactly one anchor, so it measures survival of the prepared state and is not a late-time autocorrelation); Kaplan--Meier first-exit-time distributions~\cite{kaplan1958km}; post-burn-in transition counts and dwell times; fixed late-window occupation distributions (cycles 40--49); and autocorrelation at moderate lags $k \in \{1,2,5,10\}$ with a minimum of 20 anchors per chain. A \emph{basin} is the pre-image of a local minimum under deterministic single-spin-flip steepest descent~\cite{zhu2015basin}, computed by exhaustive enumeration of the $2^{12}$ configurations.

\paragraph{Prespecified analysis plan.}
Every statistic, decision rule, interval method, burn-in rule, multiplicity treatment, and claim-to-artifact mapping in this paper was fixed in a locked analysis specification before any reported statistic was computed. We state the timing plainly: the plan was locked \emph{after} a diagnostic reanalysis of the same archived data, so approximate values of several statistics it would recompute were already known, and the specification's archived prior-knowledge disclosure enumerates them. The plan's protection against tuning-to-outcome therefore rests not on analyst blindness but on calibrated (permutation and parametric-bootstrap) thresholds, mechanical decision rules, mandatory all-cells reporting with false-discovery control, and sensitivity duplicates on its two hand-set constants. Independence units are the chain (cycled data) and the (gauge, bath, initial-state) block (endpoint data); proportions carry Clopper--Pearson 95\% intervals~\cite{clopper1934}; continuous per-chain statistics carry percentile-bootstrap intervals over chains; between-embedding variance is pooled by DerSimonian--Laird random-effects meta-analysis~\cite{dersimonian1986meta}; testing families are controlled by Benjamini--Hochberg false-discovery rate at $q \le 0.05$~\cite{benjamini1995fdr} with all cells reported. Burn-in is set per cell by a mechanical rule: $b^*$ is the earliest cycle from which $M(k)$ stays below a permutation-calibrated noise threshold through cycle 49; cells with no such $b^* \le 25$ are classified memory-retaining and excluded from pooled late-window statistics. A late-window drift test gates the vocabulary: 61 of 64 cells qualify as ``late-window quasi-stationary''; three carry ``drift detected''.

\section*{Results}

\subsection*{Instance taxonomy and schedule-dependent basin redistribution}

\begin{figure}[!htbp]
\centering
\includegraphics[width=0.98\textwidth]{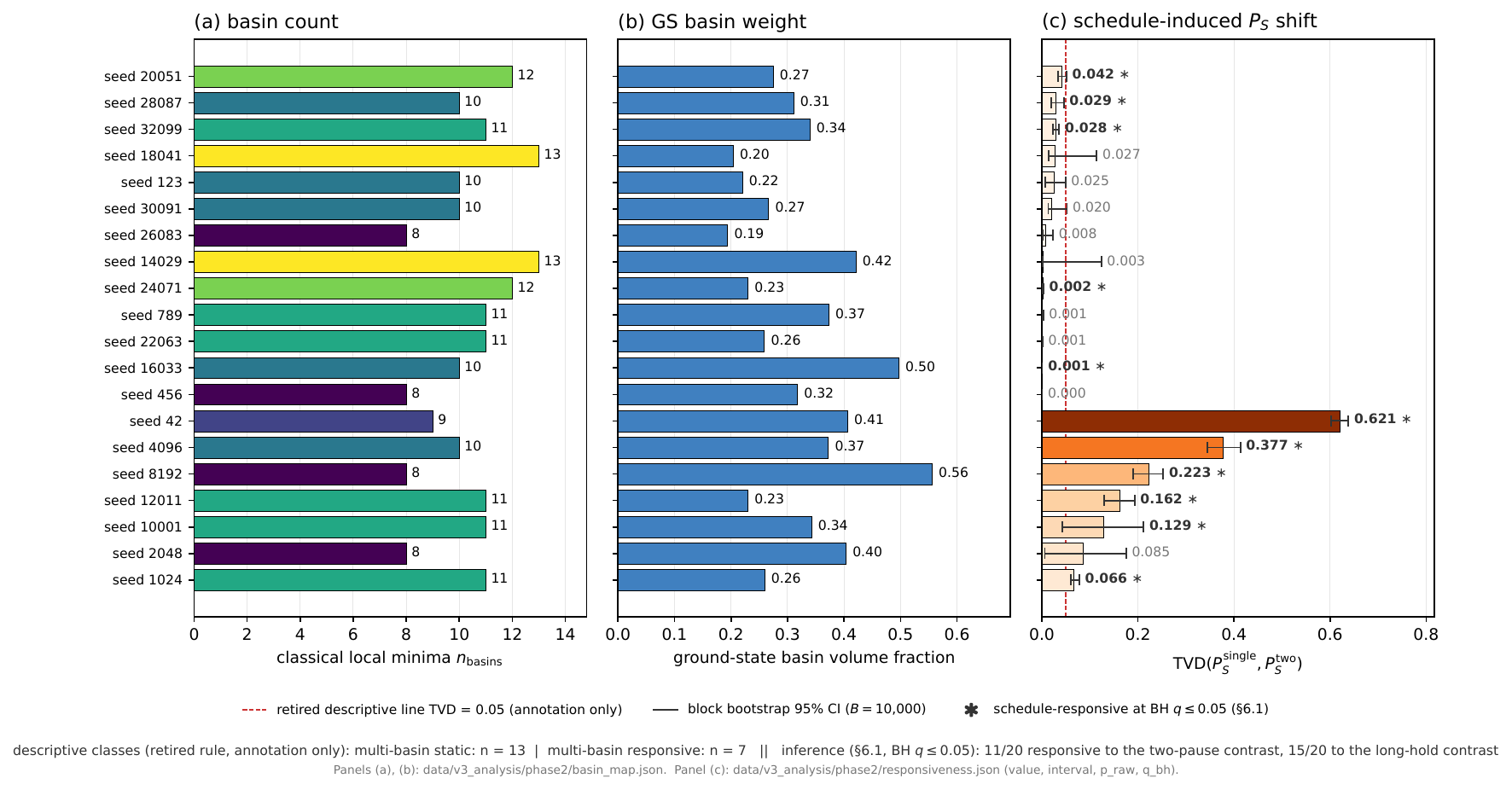}
\caption{\textbf{Instance taxonomy across the twenty training instances
(gauge-corrected).}
(a) Classical local-minimum count from exhaustive steepest-descent enumeration
of the $2^{12}$ configurations; all instances are multi-basin at the
full-system level ($n_\mathrm{basins} \in [8,13]$).
(b) Ground-state basin volume fraction; dashed line is the uniform-over-basins
reference.
(c) Schedule-induced subsystem total-variation distance between single-pause
($s_p=0.4$, 100\,$\mu$s) and two-pause ($0.45{\to}0.35$) on Advantage2,
pooled over gauges, bath states and initial states with gauge-restored
samples. The dashed 0.05 line is a descriptive annotation only: seven instances exceed it (42, 1024,
2048, 4096, 8192, 10001, 12011). Inference uses the block-permutation test
(11/20 responsive at BH $q \le 0.05$), not this line.}
\label{fig:fig2}
\end{figure}

Exhaustive basin mapping of the 20 training instances shows every instance is multi-basin at the full-system level (8--13 local minima on the $2^{12}$ hypercube; Fig.~\ref{fig:fig2}a). In the \emph{empirical} subsystem readout at the baseline schedule, three descriptive populations emerge (Table~\ref{tab:taxonomy}): seven instances are GS-dominant in readout (a post hoc concentration descriptor: one configuration carries $\ge 96\%$ of $P_S$ at baseline, the nearest non-member sits at $90\%$, and Table~\ref{tab:taxonomy} details the relation to the prespecified $\ge 88\%$ annotation), and the remaining thirteen split into seven whose readout responds to schedule change and six that are multi-basin but schedule-quiet under the descriptive total-variation measure of Fig.~\ref{fig:fig2}c.

\begin{figure}[!htbp]
\centering
\includegraphics[width=0.92\textwidth]{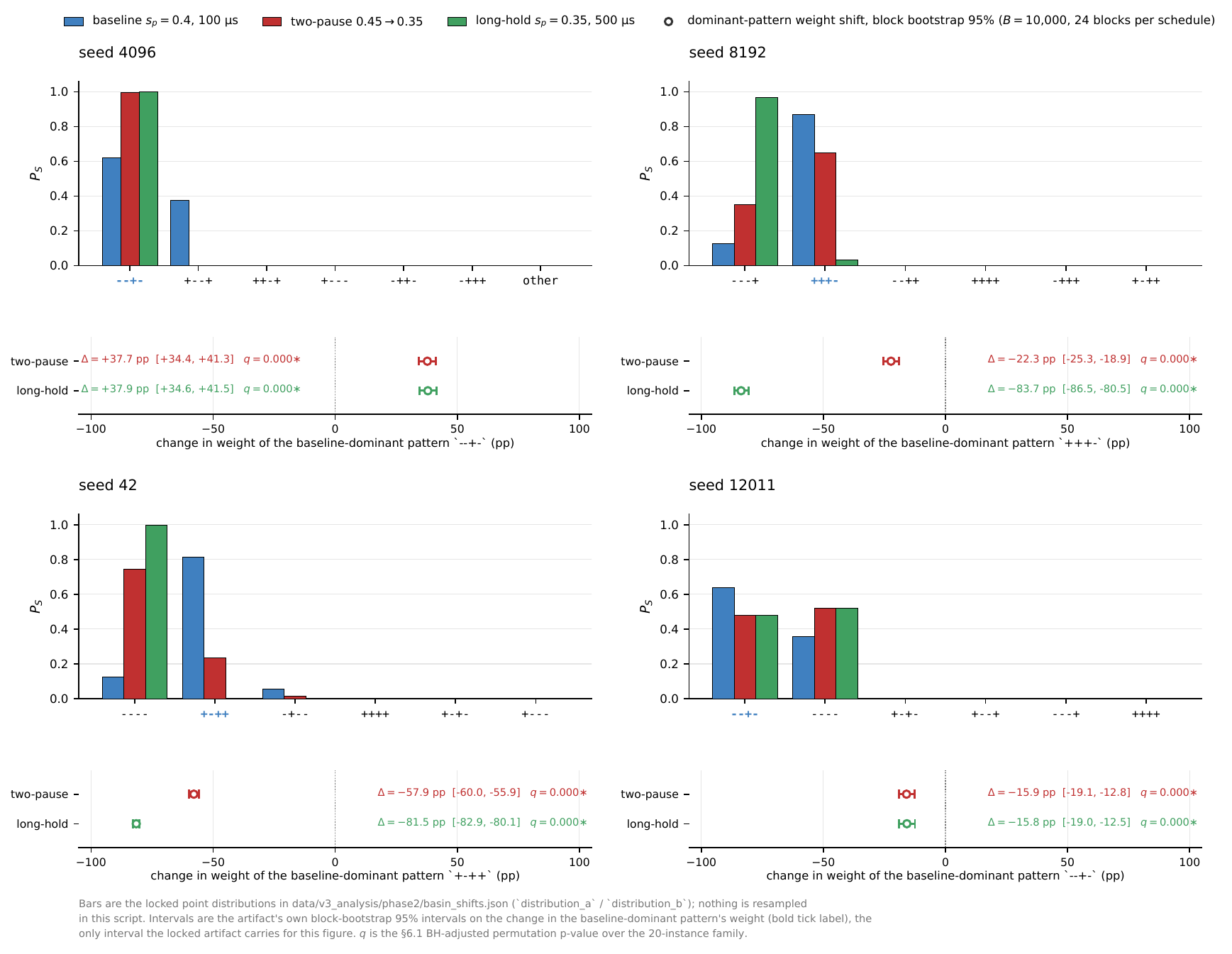}
\caption{\textbf{Schedule-dependent subsystem basin-occupation shifts on four
responsive instances (gauge-corrected).}
Empirical subsystem distribution $P_S$ under three reverse-anneal schedules:
single-pause ($s_p=0.4$, 100\,$\mu$s; blue), two-pause ($0.45 \to 0.35$; red),
and long-hold single-pause ($s_p=0.35$, 500\,$\mu$s; green), with block-bootstrap
95\% intervals over 24 (gauge, bath, initial-state) blocks per schedule.
Patterns are written in sorted-variable order.
Seed 4096: concentration onto \texttt{--+-}, $62.0\% \to 99.8\%$ (two-pause).
Seed 8192: spreading of \texttt{+++-}, $87.1\% \to 64.8\%$ (two-pause); under
long-hold the pattern \texttt{---+} grows from $12.9\%$ to $96.6\%$.
Seed 42: dominant-identity change, \texttt{+-++} at $81.6\%$ (single-pause)
$\to$ \texttt{----} at $99.8\%$ (long-hold).
Seed 12011: identity change \texttt{--+-} $63.8\% \to$ \texttt{----} $52.1\%$
(two-pause). The remaining three descriptively responsive instances (1024,
TVD $=0.07$; 2048, $0.09$; 10001, $0.13$) appear in the archived tables.}
\label{fig:fig3}
\end{figure}

Inference about schedule response does not use a fixed TVD threshold. The primary test is a block-permutation test of the schedule factor (baseline single-pause vs.\ two-pause; 24 blocks per side; $B = 10^4$), with Benjamini--Hochberg control over the 20-instance family: \textbf{11 of 20 instances are schedule-responsive} at $q \le 0.05$, and 15 of 20 respond to the long-hold contrast. Effect sizes with block-bootstrap intervals are reported for all 20 instances in the archived tables. The redistribution is large where it occurs and is not a uniform sharpening or flattening: on seed 4096 the two-pause schedule concentrates the dominant configuration \texttt{--+-} from $62.0\%$ to $99.8\%$ ($+38$\,pp); on seed 8192 it spreads \texttt{+++-} from $87.1\%$ to $64.8\%$ ($-22$\,pp); on seeds 42 and 12011 the two-pause schedule changes the \emph{identity} of the dominant configuration, and under long-hold five instances (42, 2048, 8192, 10001, 12011) change dominant identity --- seed 42 moves from \texttt{+-++} at $81.6\%$ to \texttt{----} at $99.8\%$ (Fig.~\ref{fig:fig3}). Subsystem patterns are written in sorted-variable order throughout.

\begin{table}[!htbp]
\centering
\caption{Descriptive instance taxonomy across the twenty training instances,
gauge-corrected. ``GS-dominant in readout'' = one subsystem pattern carries
$\geq 96\%$ of baseline $P_S$. This concentration descriptor is post hoc: it
is defined by a clean empirical gap
($96.0\%$ vs.\ $90.4\%$ for the nearest non-member); the prespecified
$\geq 88\%$ annotation retained in the released artifacts selects nine
instances (adding 10001 at $90.4\%$ and 14029 at $89.9\%$). No inference uses
this class. The responsive/quiet split shown here is the descriptive TVD
landscape of Fig.~\ref{fig:fig2}c; inferential responsiveness (11/20) is the
block-permutation result in the text.}
\label{tab:taxonomy}
\footnotesize
\begin{tabular}{@{}>{\raggedright\arraybackslash}p{4.1cm}>{\raggedright\arraybackslash}p{5.6cm}c>{\raggedright\arraybackslash}p{3.3cm}@{}}
\toprule
Class & Seeds & N & Observed schedule effect \\
\midrule
GS-dominant in readout                    & 456, 789, 16033, 20051, 22063, 24071, 32099 & 7 & negligible \\
Multi-basin responsive (descriptive)      & 42, 1024, 2048, 4096, 8192, 10001, 12011 & 7 & up to 38 pp shift; identity changes \\
Multi-basin quiet (descriptive)           & 123, 14029, 18041, 26083, 28087, 30091 & 6 & within block-bootstrap CI \\
\bottomrule
\end{tabular}
\end{table}

\FloatBarrier
\subsection*{Cycled dynamics: transient relaxation, then freezing}

\begin{figure}[!htbp]
\centering
\includegraphics[width=0.96\textwidth]{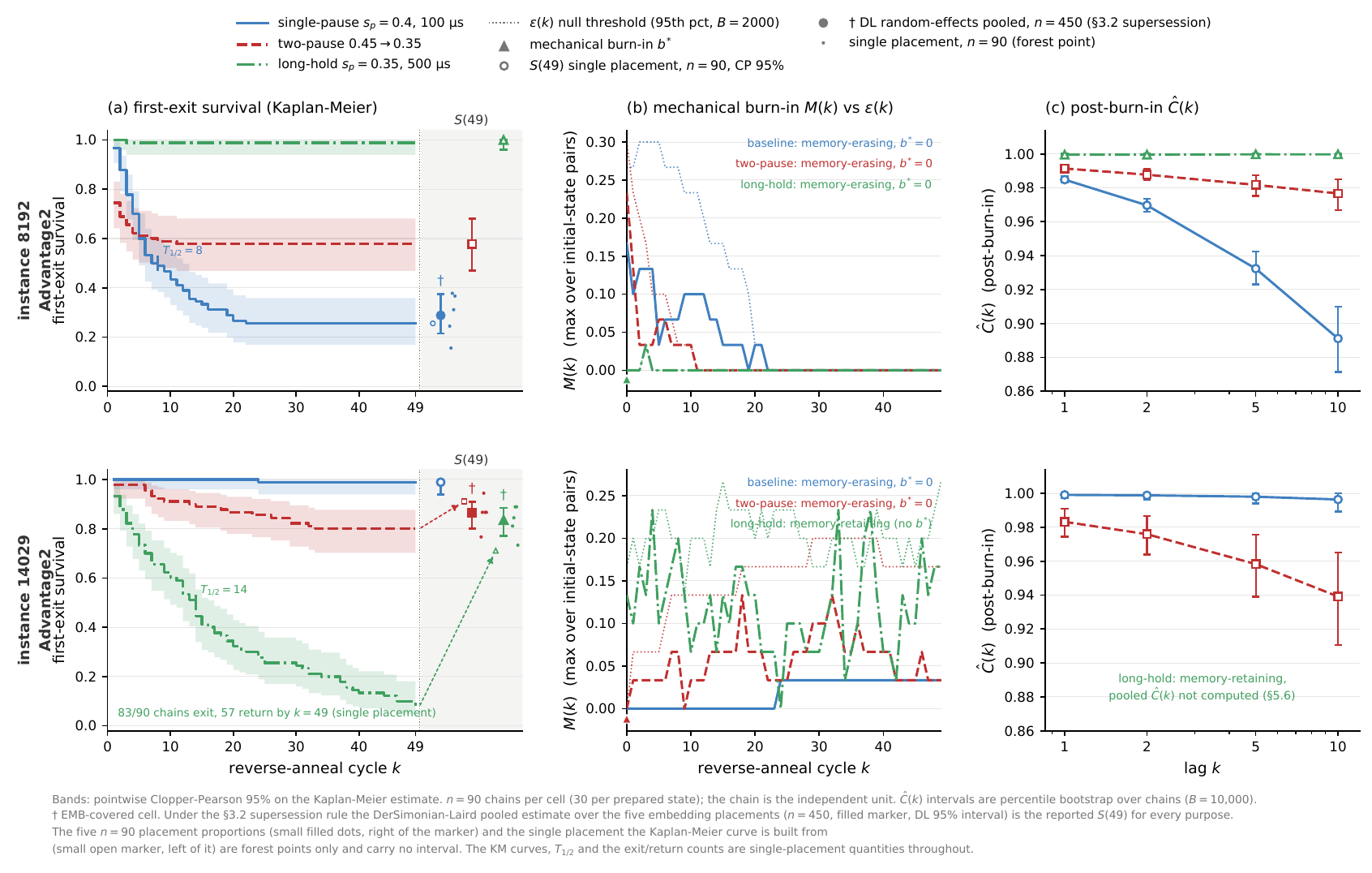}
\caption{\textbf{Cycled dynamics on Advantage2: transient, then frozen.}
Kaplan--Meier first-exit curves and moderate-lag autocorrelation with the
mechanical burn-in $b^*$ and memory classification annotated, for the six
Advantage2 cells of instances 8192 and 14029 (three schedules each).
Lag-49 quantities are preparation-survival probabilities, not
autocorrelations; error bars are Clopper--Pearson 95\% (proportions) and
chain-bootstrap 95\% (continuous statistics). For embedding-sweep-covered
cells the DerSimonian--Laird pooled estimate is primary and single
placements appear only as forest points, per the supersession rule.}
\label{fig:fig4}
\end{figure}

The cycled chains have a common structure: a transient departure from the prepared state over the first cycles, then a nearly frozen late window (Fig.~\ref{fig:fig4}). Exit and return are distinct events on these chains: on seed 14029 long-hold, only 7 of 90 chains never leave the prepared pattern (Kaplan--Meier survival at cycle 49 is $0.078$), yet 64 of 90 are back at it by cycle 49 ($S(49) = 0.711$) --- 57 chains exit and return, so first-exit curves and preparation survival carry different information and are reported separately. The mechanical burn-in rule classifies 27 of the 39 main cycled cells as memory-erasing, with realized burn-in $b^*$ between 0 and 17 cycles; 12 cells are memory-retaining and are analyzed per initial state. Single-placement preparation-survival values are: seed 8192 baseline $S(49) = 0.256$; seed 14029 long-hold $0.711$, two-pause $0.911$, baseline $0.989$. The late-window statistics show freezing, not continued motion: on seed 8192 baseline the fixed late window (cycles 40--49) is a point mass, with collision probability $1.00$, and the probability that the pattern at cycle 10 persists to cycle 49 is $0.822$ (an exploratory diagnostic, indexed as such in the release). The chain leaves its prepared state once and then freezes; nothing in these data shows continued late-time motion. Where the five-placement embedding sweep covers a cell, the random-effects pooled estimate over $n=450$ chains supersedes the single-placement value: seed 14029 long-hold pools to $S(49) = 0.835$ $[0.771, 0.884]$ (between-placement $\tau = 0.37$, $I^2 = 62\%$) against the single-placement $0.711$, seed 14029 two-pause to $0.864$ $[0.800, 0.910]$, and seed 8192 baseline to $0.288$ with placements spanning $0.156$--$0.378$. Between-placement heterogeneity is real and enters the error bars.

\FloatBarrier
\subsection*{Effective scales: the probe temperature is not the many-body scale}
\label{sec:scales}

\begin{figure}[!htbp]
\centering
\includegraphics[width=0.96\textwidth]{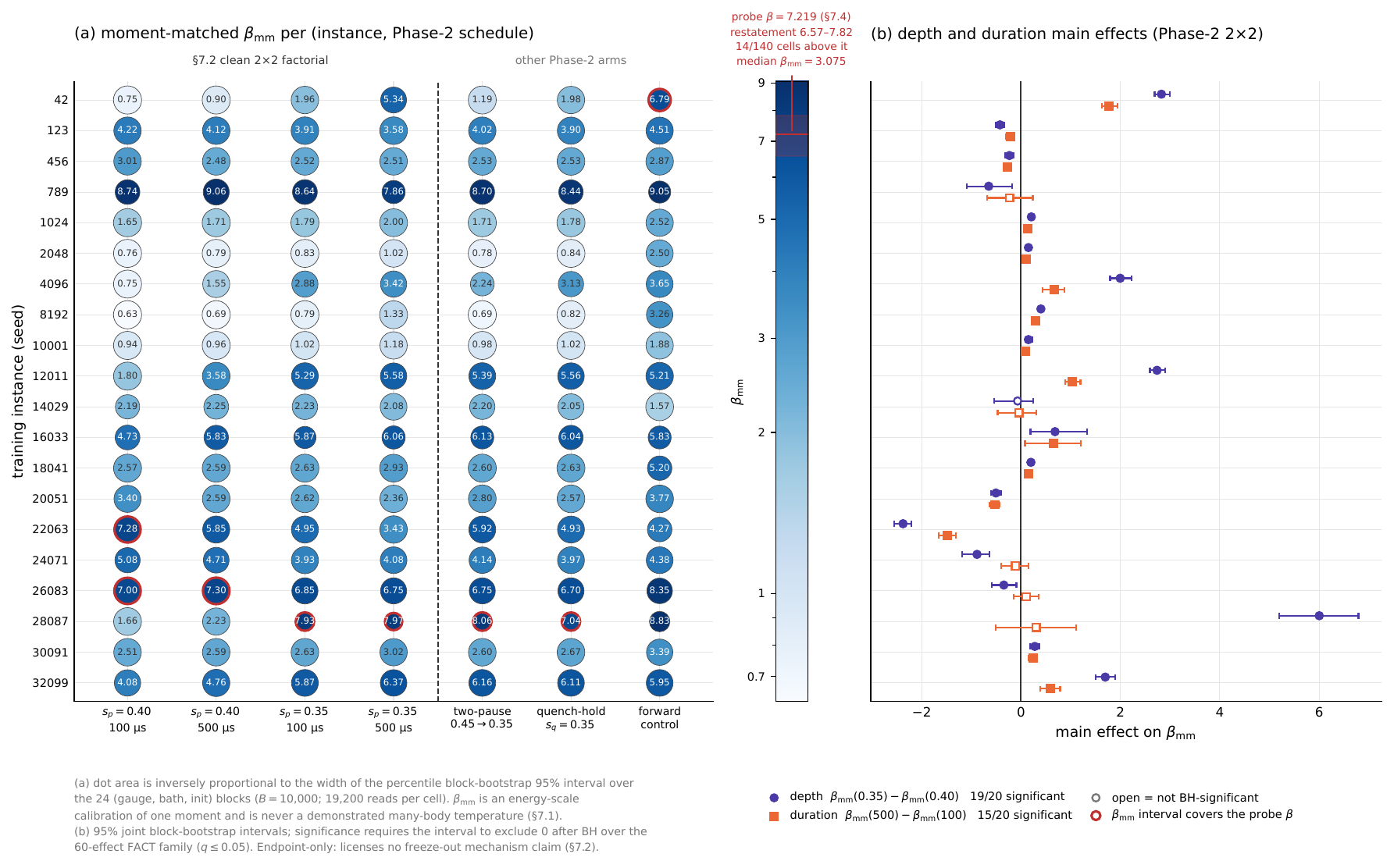}
\caption{\textbf{Effective-scale calibration.}
(a) Moment-matched inverse temperature $\beta_\mathrm{mm}$ across (instance,
schedule) cells of the endpoint atlas; the single-qubit probe value $7.22$
is marked. (b) Depth and duration main effects on $\beta_\mathrm{mm}$ from
the single-pause $2\times2$ factorial, per instance, with 95\%
block-bootstrap intervals; filled markers are BH-significant (depth 19/20,
duration 15/20).}
\label{fig:fig6}
\end{figure}

Exhaustive enumeration makes the effective scales of every cell exactly computable. Moment-matching the mean sample energy against the exact Gibbs curve gives $\beta_\mathrm{mm}$ per (instance, schedule); distribution-fitting the 16-bin subsystem marginal gives $\hat\beta$ with a parametric-bootstrap interval. Across the 140 (instance, endpoint-schedule) cells, $\beta_\mathrm{mm}$ spans $0.63$--$9.1$: the twelve-spin instances realize scales far from, and mostly much warmer than, the single-qubit probe value $7.22$ (Fig.~\ref{fig:fig6}). Two flagship cells illustrate the gap: seed 8192 baseline moment-matches to $\beta_\mathrm{mm} = 0.63$ and seed 14029 baseline to $2.19$. Even the forward-anneal control, the coldest schedule family in mean rank, has a median maximum-likelihood $\hat\beta$ of $4.1$ (total-variation fit: $3.9$), with one instance in twenty within $\pm 1$ of the probe value. Adding a transverse field to the fit family does not rescue a temperature reading: the joint $(\beta, \Gamma)$ fit returns $\Gamma^* = 0$ on the large majority of cells (103/140 endpoint cells; 45/61 main cycled cells per acquisition).

Two factors of the schedule move the scale separately. On the endpoint $2\times 2$ factorial (pause depth $\times$ pause duration), the depth main effect on $\beta_\mathrm{mm}$ is significant on 19 of 20 instances and the duration main effect on 15 of 20 (interaction: 12 of 20; BH over the 60-effect family; block-bootstrap intervals). The factorial is endpoint-only: it decomposes endpoint response and licenses no relaxation-pathway mechanism. Beyond these main effects the calibration is strongly instance-specific: the per-instance orderings of schedules by the maximum-likelihood $\hat\beta$ realize 18 distinct permutations across the 19 resolvable instances, so no universal ``deeper pause $=$ colder ensemble'' rule is supported; only the mean-rank ordering across instances is stable. The full calibration table --- $\beta_\mathrm{mm}$, $\hat\beta$, $n_\mathrm{eff}$, memory class, and drift status for every cell --- is archived and summarized in Table~\ref{tab:calibration}.

\FloatBarrier
\subsection*{Classical reference families reproduce every cell}

\begin{figure}[!htbp]
\centering
\includegraphics[width=0.98\textwidth]{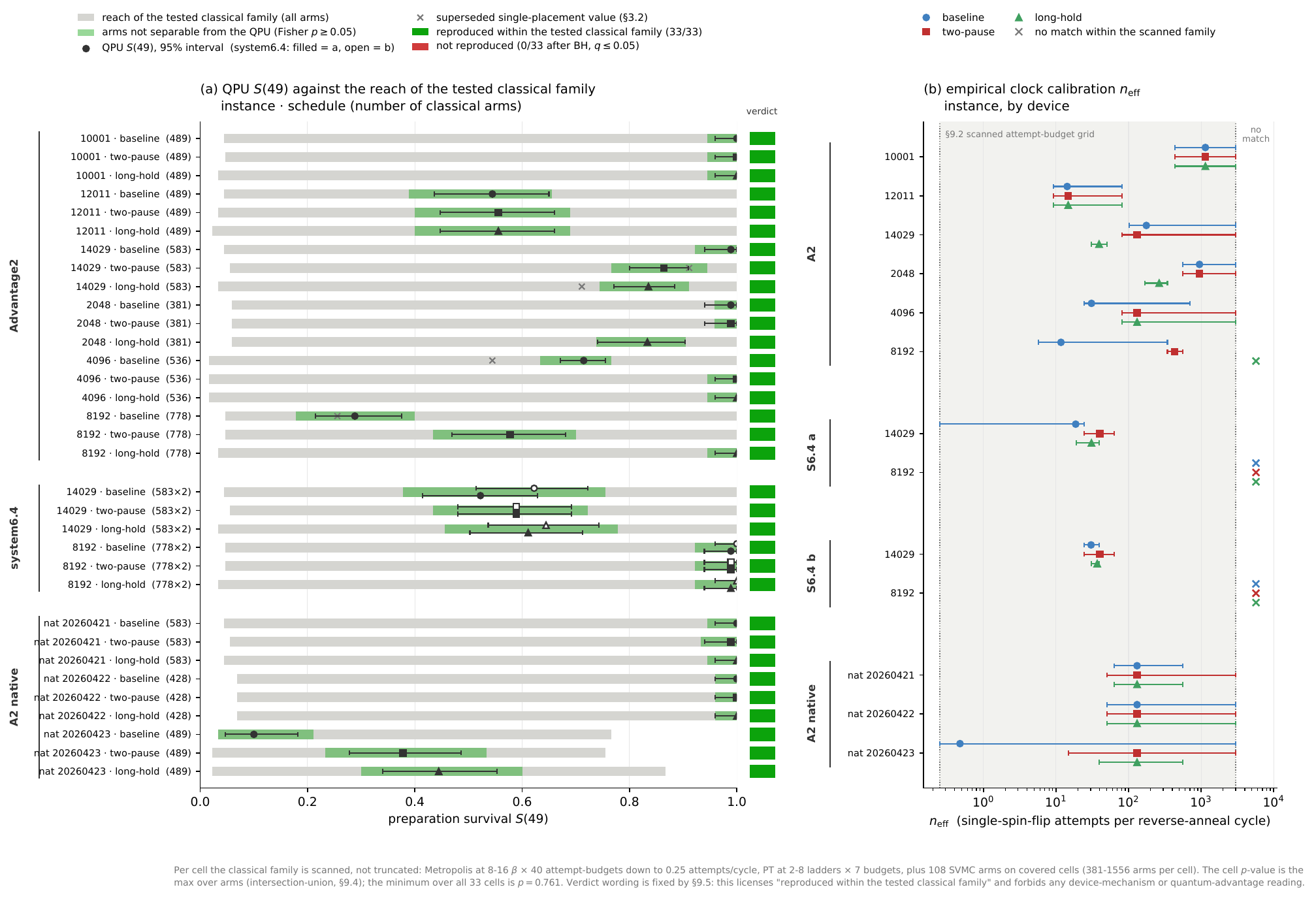}
\caption{\textbf{Classical-family comparison, all cells.}
Per (device, instance, schedule) cell: QPU preparation survival $S(49)$ with
Clopper--Pearson 95\% intervals --- except embedding-sweep-covered cells,
which show the DerSimonian--Laird pooled estimate and interval with the
superseded single-placement value as a cross --- against the reach of the
tested classical family (parallel tempering, Metropolis budget scan, and
spin-vector Monte Carlo arms), with the intersection--union verdict after
Benjamini--Hochberg control. No cell rejects (0/33; minimum $p = 0.76$). Inset: the fitted
empirical update budget $n_\mathrm{eff}$ per cell (log scale), the
``one cycle $\approx n_\mathrm{eff}$ attempts'' clock calibration.}
\label{fig:fig5}
\end{figure}

The decisive comparison sets the reference temperature per cell and treats the update budget as the continuous nuisance parameter it is. The tested classical family per cell comprises regenerated parallel tempering (per-chain random streams; ladders ending at the device-provenance and cell-fitted temperatures; budgets 1--1000 sweeps per cycle)~\cite{swendsen1986replica,hukushima1996exchange,earl2005parallel}, locally initialized single-temperature Metropolis with the attempt budget scanned over $[0.25, 3000]$ attempts per cycle, and, where available, spin-vector Monte Carlo running the actual $A(s)$, $B(s)$ schedules~\cite{shin2014svmc}. Under a per-cell intersection--union test on preparation survival with Benjamini--Hochberg control, \textbf{0 of 33 cells are outside the tested classical family}; the minimum intersection--union $p$ across all cells is $0.76$, with 381--778 arms per cell (Fig.~\ref{fig:fig5}). The low-budget end of the scan is where coarse grids mislead: plain Metropolis at the probe temperature $\beta = 7.219$ with $4.5$ attempted flips per cycle gives $S(49) = 0.263$ against the pooled QPU estimate of $0.288$ on the flagship cell (random-effects $p = 0.57$; this embedding-sweep-covered cell is compared at its pooled value throughout). We state the evidential weight of this null honestly: the tested family's total reach is broad --- spanning most of $[0,1]$ on every cell --- so coverage alone is weak evidence, and the informative content is which arms match (the statistically indistinguishable sub-band spans 3.5--62\% of arms per cell) and the $n_\mathrm{eff}$ calibration those arms define. The fitted budgets define an empirical clock calibration --- one reverse-anneal cycle $\approx n_\mathrm{eff}$ single-spin-flip attempts at the cell's fitted temperature --- which resolves within the scanned family on 32 of 39 cells. We emphasize what this calibration is not: it is a descriptive matching of statistics, not a claim of physical equality of dynamics, and it makes no statement about the presence or absence of quantum effects in the device.

The late-window occupation distributions give the same verdict with one instructive step. Against sweep-based samplers alone (PT and Metropolis), 9 of 39 per-acquisition cells reject --- concentrated on instance 14029 (7 cells) and one native-graph instance (2 cells). Admitting the schedule-aware spin-vector arms removes all nine (e.g.\ the minimum over-arm late-window TVD on seed 14029 long-hold drops from $0.098$ to $0.007$): the statistics that sweep-based classical samplers miss are within semiclassical spin-vector reach. Per the analysis plan's licensing rules, this supports ``within the tested semiclassical family'' and nothing stronger in either direction. A joint-coverage caveat is honest and archived: for four of the eight instances tested --- 8192, 14029, 12011, and the mobile native-graph instance --- no single spin-vector parameter setting reproduces all three schedules simultaneously (instances 10001, 4096, 42, and the frozen native control admit joint settings), and the spin-vector model matches the forward-anneal control far more closely (endpoint TVD $0.009$--$0.048$ across instances) than the reverse-anneal schedules; the model family covers the statistics cell by cell, not with one global clock. A post hoc robustness check, performed after the locked analyses and labeled exploratory accordingly (Addendum C of the analysis plan), finds the joint non-coverage move-set-robust: two independently gated alternative proposal kernels (local Gaussian steps of $0.5$ and $0.1$\,rad) enlarge the covered cell set --- the added arms are not inert --- yet no parameter setting of any tested kernel covers all three schedules of those four instances, and the mobile native-graph instance's best achievable marginal worsens under the local kernels; the indexed sensitivity artifact carries the per-kernel gates, coverage deltas, and its own environment record.

Cross-generation consistency follows the same pattern. The two Advantage\_system6.4 acquisitions agree with each other on every cell (Fisher $p \ge 0.23$), while the cross-\emph{device} difference on instance 14029 baseline is large ($S(49) = 0.989$ on Advantage2 vs.\ $0.522$/$0.622$ on the two S6.4 acquisitions) --- consistent with the two devices realizing different effective scales on the same instance ($\hat\beta = 3.6$ on Advantage2 vs.\ $1.4$ on the memory-erasing S6.4 acquisition; Table~\ref{tab:calibration}), and a caution against treating any single-device number as a device-independent property.

\FloatBarrier
\subsection*{Equilibrium-family residuals: condition-specific, unreplicated}

A separate, prespecified question is whether the pooled late-window \emph{distribution} of each memory-erasing cell is consistent with some member of an equilibrium family: exact classical Gibbs over $\beta$, and exact transverse-field Ising diagonal Gibbs over $(\beta, \Gamma)$ with $\Gamma$ scanned over $[0,1]$ (covering every defensible dimensionless convention for the tested pause points). Against a refit parametric-bootstrap floor at the cell's clustered effective sample size, with a materiality threshold of $0.05$ TVD and Benjamini--Hochberg control over the 27-cell family, \textbf{7 of 27 cells carry a resolved residual}: three schedules of instance 10001 on Advantage2, one cell each of instances 2048 and 8192, one native-graph cell, and instance 14029 baseline on one S6.4 acquisition (residual TVD $0.125$). Four of the seven are not equilibrium-family failures in any informative sense: on the three instance-10001 cells and the instance-2048 cell, all 90 chains are constant across the pooled window, so the empirical distribution is a point mass that no continuous equilibrium family can fit, and the memory criterion cannot flag this (all prepared states share one pattern, so $M(k) \equiv 0$). Only three resolved cells sit on mobile ensembles: instance 8192 baseline on Advantage2 (itself the one burn-in-sensitive verdict), the native-graph cell, and instance 14029 baseline on S6.4. \emph{No resolved residual meets the prespecified replication requirement} (both S6.4 acquisitions, or an Advantage2 cell plus a same-instance S6.4 cell) --- and for five of the seven the requirement was unsatisfiable by construction, because no second-device acquisition exists for those instances; for instance 14029 the second acquisition's cell is memory-retaining and never enters the family, and the single cell actually tested for replication (instance 8192 baseline) failed it. These residuals are therefore reported as secondary, unreplicated observations and deliberately carry no abstract-level claim. Hamiltonian-misspecification cover for them is model-dependent: an i.i.d.\ perturbation ensemble at the intrinsic-control-error scale ($\delta J \sim \mathcal{N}(0, 0.015)$, $\delta h \sim \mathcal{N}(0, 0.02)$) does not cover the S6.4 instance-14029 residual, while a background-susceptibility-correlated variant does --- so a correlated programming-bias origin cannot be excluded, which the unreplicated status already reflects. Sensitivity duplicates (burn-in offsets, a fixed late window, materiality thresholds $0.03/0.05/0.075$) move the resolved count between 5 and 9 and change one cell's verdict (seed 8192 baseline, flagged burn-in-sensitive); the qualitative conclusion --- condition-specific, concentrated, unreplicated --- is stable (Fig.~\ref{fig:fig7}).

\FloatBarrier
\subsection*{Predictor failure and pilot reconciliation under corrected conditioning}
\label{sec:conditioning}

A naive endpoint reference conditions the classical and quantum targets on the \emph{initial} environment configuration, as if the environment were held fixed. It is initialized only --- no field pins it --- and on several instances it leaves the initial configuration in most cycles; on 236 of 1350 conditions the initial-conditioned quantum reference conditions on an event to which its own model assigns probability below $10^{-12}$. The corrected references use the observed final-environment mixture at the cell's own $\beta_\mathrm{mm}$, with the quantum reference at the schedule's dimensionless $\Gamma = A(s_p)/B(s_p)$. The correction is consequential: the sign of $\Delta D = D_\mathrm{cl} - D_q$ flips on roughly $44\%$ of conditions, and the initial-conditioned variant is preserved in the artifacts as provenance.

Two analyses are restated under the corrected targets. First, the pre-registered linear predictor of schedule sensitivity from landscape features: applied unchanged to the corrected target, no model passes either pre-registered criterion (held-out Pearson $p < 0.05$; held-out $R^2 > 0$), and the initial-conditioned execution also fails both. Because the target changed, this is a corrected-target re-analysis, no longer a clean execution of the original pre-registration; the strongest corrected correlation (mean-barrier, $r = -0.61$, $p = 0.061$) is opposite in sign to the pre-registered direction and carries $R^2 = -3.0$. Second, the companion pilot's reported enhancement --- a two-pause schedule doubling the fraction of instances whose readout prefers the quantum target ($17\% \to 44\%$ on 40 instances)~\cite{locth1_companion} --- is contradicted, not merely unsupported: under corrected conditioning at instance level the two-pause inversion rate is $1/20$ (Clopper--Pearson $[0.00, 0.25]$), significantly \emph{below} the pilot rate (Fisher $p = 0.0013$), and no schedule ordering survives among rates of $0$--$1$ instance in twenty across all seven schedule families. The pilot's apparent doubling is attributable to the combination of small-sample noise ($\pm 15$\,pp at $n=40$) and initial-environment conditioning evaluated at the probe temperature.

\FloatBarrier
\section*{Discussion}

We set out to test whether reverse-anneal schedule shape controls subsystem sampling beyond a temperature rescaling, and at what scales classical processes reproduce the observed cycled dynamics. Four findings, in the order the evidence supports them.

\textbf{First}, schedule shape is a real, instance-specific control on basin occupation: 11 of 20 instances respond under a calibrated test. Statistical response and magnitude are distinct: where the effect is large it is very large (weight shifts up to 38 percentage points; changes of dominant-configuration identity), while several statistically responsive instances shift by far less, and the release tables report significance and effect size separately for every instance. The effect is not a uniform sharpen-or-flatten knob; it reassigns probability among competing near-degenerate minima in a way that depends jointly on the instance and the schedule.

\textbf{Second}, the effective scales that a schedule realizes are properties of the (instance, schedule, device) triple, not of the device alone. The single-qubit probe temperature ($\beta \approx 7.2$) does not transfer as a universal many-body scale ($\beta_\mathrm{mm}$ spans $0.63$--$9.1$ across cells, mostly far below the probe value); pause depth and pause duration each move the scale on most instances; and the same instance realizes $\hat\beta = 3.6$ on one device generation and $1.4$ on the other. Any application that treats the reverse-anneal schedule as a temperature knob needs this calibration per instance and per schedule; the archived table provides it for this working point.

\textbf{Third}, nothing in this dataset requires more than the tested classical families. With the update budget scanned continuously --- including the sub-sweep regime --- every cell's preparation survival is reproduced within the tested classical family, and the late-window statistics that sweep-based samplers miss are within reach of a spin-vector model running the actual schedules. We state the licensing boundary explicitly: these results do not show that the device \emph{is} classical, and the empirical clock calibration $n_\mathrm{eff}$ is a matching of statistics, not an identification of dynamics; symmetric restraint applies in both directions.

\textbf{Fourth}, the corrected negative results carry information. The pre-registered structural predictor fails out-of-sample under both conditionings, locating schedule sensitivity at a level of basin geometry not captured by the tested scalar landscape moments. The companion pilot's two-pause enhancement is contradicted at $p = 0.0013$ under corrected conditioning: on this working point, no tested schedule preferentially aligns readout with the quantum target once references are evaluated at the scales the data actually realize.

The equilibrium-family residuals on mobile ensembles --- three cells, each resolved on a single acquisition, one of them burn-in-sensitive --- are the one open observation, and they are deliberately unclaimed here. If processor access permits, the decisive experiment is already specified: reacquire instances 8192 and 14029 on both device generations in a memory-erasing regime and apply the archived decision rule unchanged.

Beyond the corrected claims, the practical deliverable of this study is methodological: a fully prespecified, artifact-traced analysis pattern for cycled-anneal experiments --- mechanical burn-in rules, exact-enumeration references wherever the instance size permits, continuous nuisance-parameter scans for classical baselines, replication requirements for anomaly claims, and licensing tables that fix in advance what each comparison may and may not conclude. At $N = 12$ every one of these is computable exactly; we suggest that no claim of classically unmatched behavior on exhaustively enumerable instances should be published without them.

All claims here are bounded by the working point: twelve-qubit instances of one mixed-frustration family at one $(\lambda, W, p_S)$ setting, three cycled schedules, two device generations, and no size scaling --- no claim generalizes beyond this scope, and the schedule-response fractions carry no known size dependence.

\FloatBarrier
\section*{Methods}

\paragraph{Hardware, acquisitions, and identity.}
Advantage2 (\texttt{Advantage2\_system1}, Zephyr) and Advantage\_system6.4 (Pegasus), accessed through D-Wave Leap; solver properties and anneal-schedule tables per the solver documentation~\cite{dwave_system_docs}, with the $A(s)/B(s)$ table archived with checksum provenance. Acquisition dates: endpoint atlas 2026-04-16/17; $\beta$ calibration and cycled acquisitions 2026-04-18; family controls and held-out validation 2026-04-19; embedding sweep and cross-solver acquisitions 2026-04-20; native-graph control 2026-04-21. The solver did not expose a calibration epoch; runs spanning six days may cross calibration cycles, and between-run variation is treated as uncertainty where the design permits (random-effects pooling; per-acquisition analysis). Device identity of every run is taken from its manifest, never from directory names. The cycled Advantage2 acquisition terminated during its seventh instance; the six complete instances constitute the per-cell families, the truncation point was set by run order and wall time (outcome-independent), and the affected instance is reported descriptively only.

\paragraph{Embedding and programming scale.}
Endpoint acquisitions used one cached minor embedding per instance~\cite{cai2014minor}, fixed across schedules, gauges, bath and initial states; sixteen instances embedded at chain length 1 and four (4096, 14029, 18041, 28087) carry a single length-2 chain with Ocean-default chain strength and majority-vote resolution (chain-break fractions were not recorded for these runs)~\cite{marshall2020embedding}. The cycled acquisition used an independently seeded embedding per instance and recorded no embedding metadata; the five-placement sweep bounds placement variation and enters the analysis as a random effect. The cross-solver acquisitions used length-1 embeddings on all twelve logical qubits with recorded chain-break fraction identically zero; the \texttt{chain\_strength} parameter is inoperative at chain length 1. All instances have $\max|h| \le 0.993$ and $\max|J| = 1.000$, so automatic scaling is a no-op for every Hamiltonian in this work regardless of the setting used; the $\beta$ probes and the native-graph control set \texttt{auto\_scale=False} explicitly. Gauge averaging was applied in the endpoint atlas (four gauges; samples analyzed in the restored frame throughout) and not in the cycled acquisition (single gauge); the five-placement sweep controls qubit-specific defects but not the sign structure of programming errors, which is instead addressed by the perturbed-Hamiltonian ensembles.

\paragraph{Native-graph control.}
Three random 12-node connected induced subgraphs of the Advantage2 hardware graph, Hamiltonians defined directly on native physical edges, queried through the bare sampler with \texttt{auto\_scale=False} and no minor embedding; one instance drew zero frustrated $S$--$S$ bonds and serves as the designed unfrustrated control. Native cells enter every family under the identical rules.

\paragraph{Exact references.}
Every instance is exhaustively enumerable: classical Gibbs quantities from the $4096$-state enumeration; transverse-field Ising diagonal Gibbs from one eigendecomposition of $H_\mathrm{cl} - \Gamma \sum_i \sigma^x_i$ per (instance, $\Gamma$), $\Gamma \in [0,1]$ in steps of $0.02$. The moment-match $\beta_\mathrm{mm}$ solves $\langle E \rangle_\beta = \bar E_\mathrm{obs}$ on the exact curve (unique root by monotonicity); $\hat\beta$ minimizes the total-variation distance of the 16-bin subsystem marginal with a parametric-bootstrap interval; the residual floor is a refit parametric bootstrap at the cell's design-effect-adjusted sample size. Regenerated parallel tempering is validated against exact enumeration (long-budget pooled windows within the parametric floor of exact Gibbs; replica-exchange acceptance 0.48--0.999); Metropolis is validated by detailed balance ($<3\times10^{-14}$ relative violation) and exact-Gibbs agreement where equilibration is reachable. The spin-vector model is validated in the hard-spin invariance sector, against exact rotor quadrature at $A \to 0$, and at static $(A, B)$ working points, before any device comparison.

\paragraph{Statistics.}
As prespecified: Clopper--Pearson intervals for proportions (bootstrap intervals degenerate to zero width on frozen cells and are not used); percentile bootstrap over chains for continuous statistics; DerSimonian--Laird random effects over embedding placements; block permutation for endpoint schedule response; intersection--union tests for composite ``no tested arm reproduces'' nulls; Benjamini--Hochberg control within each prespecified family with all cells reported. Randomness derives from a single master seed through documented spawn paths with one stream per chain, bootstrap, and permutation. The analysis specification, its prior-knowledge disclosure, every artifact with its producing code reference, and a machine-readable index mapping each reported number to its artifact are part of the data release.

\paragraph{Data and code availability.}
The release archive contains: raw sample records for every acquisition (NPZ samplesets of samples, energies, and variables), each with a per-acquisition conditions sidecar carrying per-condition metadata --- the solver identifier from the manifest, the solver timing block, and the schedule waypoints --- and a manifest carrying solver identity and schedule definitions (embedding metadata where recorded: the cycled acquisition recorded none; chain-break fractions only on the cross-solver runs, identically zero); frozen instance definitions as explicit $(h, J, S, E)$ dictionaries with energy-vector checksums; the locked analysis specification with its addenda; all regenerated classical-baseline records with per-chain seed derivations; every derived artifact with code references; and the reproduction scripts. Superseded earlier baseline records are preserved in the archive as provenance, labeled with their defects, and are used for no inference. The archive will be deposited with a DOI at acceptance; a reviewer-access copy is available on request during peer review.

\paragraph{LLM disclosure.}
LLM-based coding assistants were used for code scaffolding, for internal reanalysis and verification, and for language editing. All scientific decisions, code validation, data analysis, and interpretation were performed and verified by the author.

\section*{Acknowledgments}

The author thanks the D-Wave team for access to quantum processing units through the D-Wave Leap platform.

\bibliographystyle{unsrt}
\bibliography{refs}

\clearpage
\appendix
\section*{Appendix A: calibration table and residual verdicts}

\begin{table}[!htbp]
\centering
\caption{Effective-scale calibration summary for the cycled cells of the two
two flagship instances on both devices: moment-matched
$\beta_\mathrm{mm}$, distribution-fit $\hat\beta$, empirical update budget
$n_\mathrm{eff}$ (attempts per cycle; ``---'' = no match within the scanned
family), memory class, and window status (q-stat.\ = late-window
quasi-stationary per the drift test). $\beta_\mathrm{mm}$ derives from the
Advantage2 endpoint atlas for the corresponding schedule and is therefore
identical across device rows by construction; device-specific scales are
carried by $\hat\beta$. $\hat\beta$ intervals reaching $12.00$ are censored
at the fit-grid edge. The complete 39-cell table with intervals is
Table~SI (appendix), machine-readable in the release.}
\label{tab:calibration}
\input{tables/calibration_summary}
\end{table}

\begin{figure}[!htbp]
\centering
\includegraphics[width=0.96\textwidth]{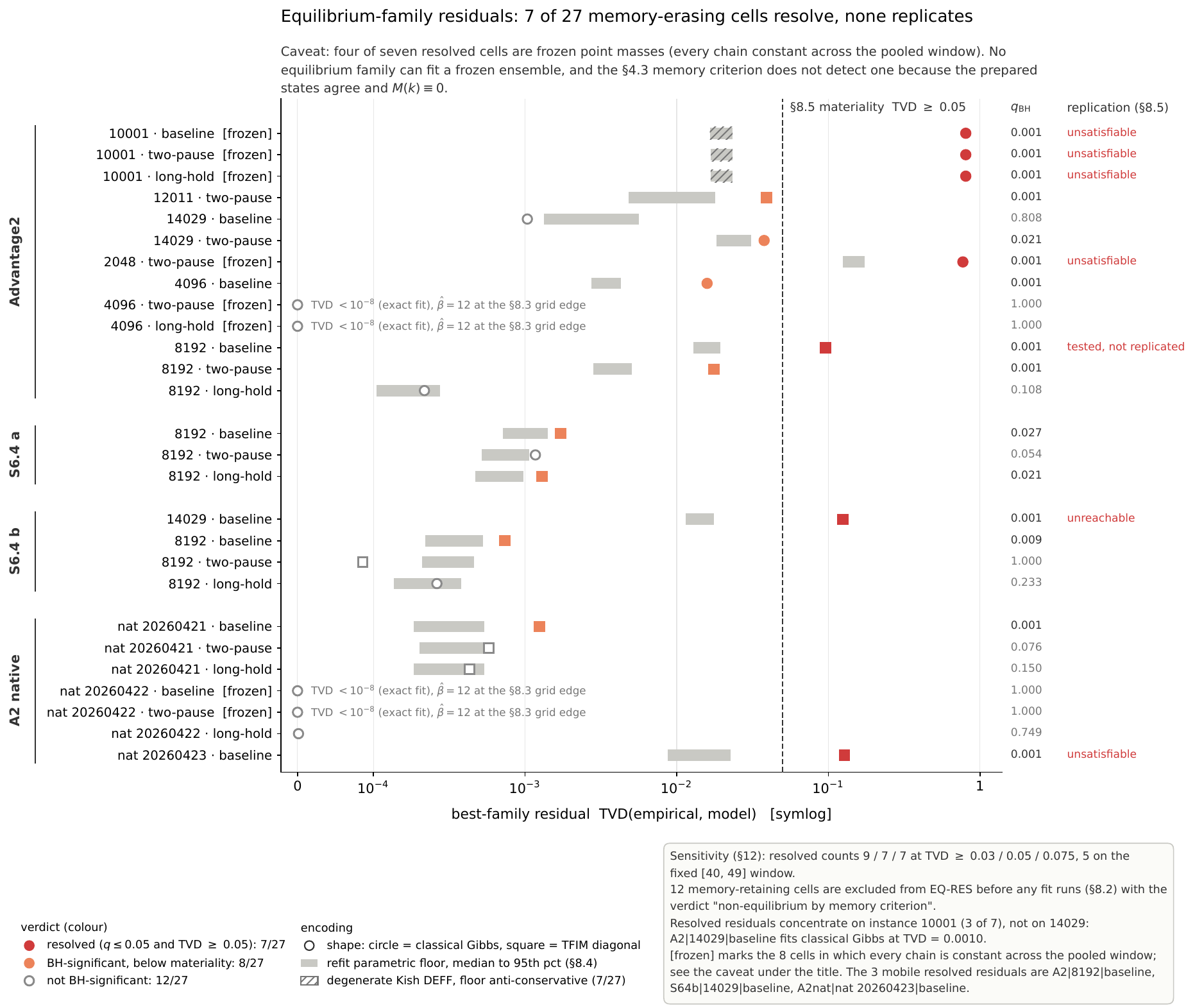}
\caption{\textbf{Equilibrium-family residual verdicts, all memory-erasing
cells.} Best-family residual total-variation distance per cell against its
refit parametric-bootstrap floor, with the prespecified materiality threshold
(0.05) marked; resolved cells (7/27) are highlighted with their replication
status, and frozen point-mass ensembles (all 90 chains constant across the
pooled window; no continuous equilibrium family can fit them) are tagged.
No resolved residual meets the replication requirement; for five of seven it
was unsatisfiable by construction, and the one tested cell failed it.}
\label{fig:fig7}
\end{figure}

\FloatBarrier
\section*{Appendix B: conclusion-licensing and all-cells tables}

Per the locked analysis plan, the licensing table fixes in advance what each
comparator may and may not conclude, and every testing family is reported in
full --- estimate, interval, raw $p$, and Benjamini--Hochberg $q$ for every
cell, with no selective reporting.

\input{tables/licensing_table}
\input{tables/calibration_full}
\input{tables/family_p2resp}
\input{tables/family_p2resp_lh}
\input{tables/family_fact}
\input{tables/family_eqres}
\input{tables/family_dynclass}

\end{document}

%% file: tables/calibration_summary.tex
\footnotesize
\begin{tabular}{llllllll}
\toprule
Device & Instance & Schedule & $\beta_\mathrm{mm}$ & $\hat\beta$ [95\% CI] & $n_\mathrm{eff}$ & Memory & Window \\
\midrule
Advantage2 & 8192 & single-pause (0.40, 100\,$\mu$s) & 0.63 & 0.91 [0.87, 0.96] & 12 & erasing & q-stat. \\
Advantage2 & 8192 & long-hold (0.35, 500\,$\mu$s) & 1.33 & 2.27 [2.01, 12.00] & --- & erasing & q-stat. \\
Advantage2 & 8192 & two-pause (0.45$\to$0.35) & 0.69 & 1.27 [1.19, 1.37] & 432 & erasing & q-stat. \\
Advantage2 & 14029 & single-pause (0.40, 100\,$\mu$s) & 2.19 & 3.56 [2.96, 12.00] & 176 & erasing & q-stat. \\
Advantage2 & 14029 & long-hold (0.35, 500\,$\mu$s) & 2.08 & --- & 39 & retaining & q-stat. \\
Advantage2 & 14029 & two-pause (0.45$\to$0.35) & 2.20 & 2.08 [1.85, 2.39] & 131 & erasing & q-stat. \\
S6.4 (acq.~a) & 8192 & single-pause (0.40, 100\,$\mu$s) & 0.63 & 1.54 [1.46, 1.66] & --- & erasing & q-stat. \\
S6.4 (acq.~a) & 8192 & long-hold (0.35, 500\,$\mu$s) & 1.33 & 1.67 [1.56, 1.86] & --- & erasing & q-stat. \\
S6.4 (acq.~a) & 8192 & two-pause (0.45$\to$0.35) & 0.69 & 1.66 [1.56, 1.83] & --- & erasing & q-stat. \\
S6.4 (acq.~a) & 14029 & single-pause (0.40, 100\,$\mu$s) & 2.19 & --- & 19 & retaining & q-stat. \\
S6.4 (acq.~a) & 14029 & long-hold (0.35, 500\,$\mu$s) & 2.08 & --- & 31 & retaining & q-stat. \\
S6.4 (acq.~a) & 14029 & two-pause (0.45$\to$0.35) & 2.20 & --- & 40 & retaining & q-stat. \\
S6.4 (acq.~b) & 8192 & single-pause (0.40, 100\,$\mu$s) & 0.63 & 1.95 [1.78, 2.27] & --- & erasing & q-stat. \\
S6.4 (acq.~b) & 8192 & long-hold (0.35, 500\,$\mu$s) & 1.33 & 2.10 [1.89, 12.00] & --- & erasing & q-stat. \\
S6.4 (acq.~b) & 8192 & two-pause (0.45$\to$0.35) & 0.69 & 2.01 [1.82, 12.00] & --- & erasing & q-stat. \\
S6.4 (acq.~b) & 14029 & single-pause (0.40, 100\,$\mu$s) & 2.19 & 1.43 [1.39, 1.47] & 30 & erasing & q-stat. \\
S6.4 (acq.~b) & 14029 & long-hold (0.35, 500\,$\mu$s) & 2.08 & --- & 37 & retaining & drift \\
S6.4 (acq.~b) & 14029 & two-pause (0.45$\to$0.35) & 2.20 & --- & 40 & retaining & q-stat. \\
\bottomrule
\end{tabular}

%% file: tables/licensing_table.tex
\begin{table*}[htbp]
\centering
\caption{Conclusion licensing. Transcribed verbatim from the pre-registered analysis specification (\textsc{analysis\_spec} \S9.5): what each comparator in this study does, and does not, license as a conclusion.}
\label{tab:licensing}
\footnotesize
\begin{tabular}{lll}
\toprule
\parbox[t]{0.15\textwidth}{\raggedright Comparator} & \parbox[t]{0.40\textwidth}{\raggedright May conclude} & \parbox[t]{0.36\textwidth}{\raggedright May NOT conclude} \\
\midrule
\parbox[t]{0.15\textwidth}{\raggedright Exact enumeration} & \parbox[t]{0.40\textwidth}{\raggedright equilibrium values, $C_\mathrm{eq}$, Gibbs/TFIM marginals} & \parbox[t]{0.36\textwidth}{\raggedright anything about dynamics} \\
\parbox[t]{0.15\textwidth}{\raggedright PT (regenerated)} & \parbox[t]{0.40\textwidth}{\raggedright equilibrium consistency; strongest-mixing classical envelope at tested budgets} & \parbox[t]{0.36\textwidth}{\raggedright timescale claims (PT sweeps have no physical clock); ``no classical process can match''} \\
\parbox[t]{0.15\textwidth}{\raggedright Metropolis $n_\mathrm{eff}$ fit} & \parbox[t]{0.40\textwidth}{\raggedright ``one cycle $\approx$ $n_\mathrm{eff}$ attempts at $\beta$'' as an \textbf{empirical clock calibration}; budget ordering across schedules} & \parbox[t]{0.36\textwidth}{\raggedright physical equality of dynamics; microscopic mechanism; presence or absence of quantum effects} \\
\parbox[t]{0.15\textwidth}{\raggedright SVMC (schedules)} & \parbox[t]{0.40\textwidth}{\raggedright match $\to$ observed schedule dependence is within semiclassical spin-vector reach (no coherence language anywhere); mismatch $\to$ ``outside the tested SVMC family''} & \parbox[t]{0.36\textwidth}{\raggedright match $\to$ ``the device is classical''; mismatch $\to$ ``quantum'' or ``coherent''} \\
\parbox[t]{0.15\textwidth}{\raggedright DYN-CLASS verdict} & \parbox[t]{0.40\textwidth}{\raggedright ``not reproduced within the tested classical family''} & \parbox[t]{0.36\textwidth}{\raggedright quantum advantage, quantum pathway, or any device-mechanism attribution} \\
\bottomrule
\end{tabular}
\vspace{1ex}
\begin{minipage}{\linewidth}
\scriptsize
\textit{Notes.} Row text is the pre-registered wording; the only edits are markdown emphasis rendered as \LaTeX{} bold, straight quotes rendered as typographic quotes, dashes, and the subscripted identifiers $C_\mathrm{eq}$, $n_\mathrm{eff}$, $\beta$.
\end{minipage}
\end{table*}

%% file: tables/calibration_full.tex
\begin{table*}[htbp]
\centering
\caption{Per-cell effective-scale calibration, all cells. Moment-matched inverse temperature $\beta_\mathrm{mm}$, equilibrium-fit $\hat\beta$, effective update budget $n_\mathrm{eff}$, memory class and late-window status for every Phase-3A cell (39 cells).}
\label{tab:calibration_full}
\tiny
\setlength{\tabcolsep}{3pt}
\begin{tabular}{llllllll}
\toprule
Device & Instance & Schedule & $\beta_\mathrm{mm}$ [95\% CI] & $\hat\beta$ [95\% CI] & $n_\mathrm{eff}$ [95\% CI] & Memory & Window \\
\midrule
Advantage2 & 10001 & single-pause (0.40, 100\,$\mu$s) & 0.94 [0.92, 0.96] & 0.62 [0.61, 0.63] & 1145 [437, 3000] & erasing & q-stat. \\
Advantage2 & 10001 & long-hold (0.35, 500\,$\mu$s) & 1.18 [1.09, 1.33] & 0.62 [0.61, 0.63] & 1145 [437, 3000] & erasing & q-stat. \\
Advantage2 & 10001 & two-pause (0.45$\to$0.35) & 0.98 [0.96, 1.00] & 0.62 [0.61, 0.63] & 1145 [437, 3000] & erasing & q-stat. \\
Advantage2 & 12011 & single-pause (0.40, 100\,$\mu$s) & 1.80 [1.79, 1.82] & --- & 14 [9, 81] & retaining & q-stat. \\
Advantage2 & 12011 & long-hold (0.35, 500\,$\mu$s) & 5.58 [5.37, 5.83] & --- & 15 [9, 81] & retaining & q-stat. \\
Advantage2 & 12011 & two-pause (0.45$\to$0.35) & 5.39 [5.23, 5.57] & 5.61 [5.27, 12.00] & 15 [9, 81] & erasing & q-stat. \\
Advantage2 & 14029 & single-pause (0.40, 100\,$\mu$s) & 2.19 [1.88, 2.87] & 3.56 [2.96, 12.00] & 176 [103, 3000] & erasing & q-stat. \\
Advantage2 & 14029 & long-hold (0.35, 500\,$\mu$s) & 2.08 [1.91, 2.28] & --- & 39 [31, 50] & retaining & q-stat. \\
Advantage2 & 14029 & two-pause (0.45$\to$0.35) & 2.20 [1.93, 2.67] & 2.08 [1.85, 2.39] & 131 [81, 3000] & erasing & q-stat. \\
Advantage2 & 2048 & single-pause (0.40, 100\,$\mu$s) & 0.76 [0.74, 0.78] & --- & 949 [556, 3000] & retaining & q-stat. \\
Advantage2 & 2048 & long-hold (0.35, 500\,$\mu$s) & 1.02 [0.94, 1.11] & --- & 264 [167, 343] & retaining & q-stat. \\
Advantage2 & 2048 & two-pause (0.45$\to$0.35) & 0.78 [0.77, 0.80] & 0.42 [0.30, 0.52] & 949 [556, 3000] & erasing & q-stat. \\
Advantage2 & 4096 & single-pause (0.40, 100\,$\mu$s) & 0.75 [0.73, 0.77] & 1.82 [1.74, 1.92] & 31 [24, 707] & erasing & q-stat. \\
Advantage2 & 4096 & long-hold (0.35, 500\,$\mu$s) & 3.42 [3.13, 3.65] & 12.00 [12.00, 12.00] & 131 [81, 3000] & erasing & q-stat. \\
Advantage2 & 4096 & two-pause (0.45$\to$0.35) & 2.24 [1.94, 3.05] & 12.00 [12.00, 12.00] & 131 [81, 3000] & erasing & q-stat. \\
Advantage2 & 8192 & single-pause (0.40, 100\,$\mu$s) & 0.63 [0.62, 0.64] & 0.91 [0.87, 0.96] & 12 [6, 343] & erasing & q-stat. \\
Advantage2 & 8192 & long-hold (0.35, 500\,$\mu$s) & 1.33 [1.30, 1.37] & 2.27 [2.01, 12.00] & --- & erasing & q-stat. \\
Advantage2 & 8192 & two-pause (0.45$\to$0.35) & 0.69 [0.69, 0.70] & 1.27 [1.19, 1.37] & 432 [343, 556] & erasing & q-stat. \\
\midrule
S6.4 (acq.~a) & 14029 & single-pause (0.40, 100\,$\mu$s) & 2.19 [1.88, 2.87] & --- & 19 [0, 24] & retaining & q-stat. \\
S6.4 (acq.~a) & 14029 & long-hold (0.35, 500\,$\mu$s) & 2.08 [1.91, 2.28] & --- & 31 [19, 39] & retaining & q-stat. \\
S6.4 (acq.~a) & 14029 & two-pause (0.45$\to$0.35) & 2.20 [1.93, 2.67] & --- & 40 [24, 64] & retaining & q-stat. \\
S6.4 (acq.~a) & 8192 & single-pause (0.40, 100\,$\mu$s) & 0.63 [0.62, 0.64] & 1.54 [1.46, 1.66] & --- & erasing & q-stat. \\
S6.4 (acq.~a) & 8192 & long-hold (0.35, 500\,$\mu$s) & 1.33 [1.30, 1.37] & 1.67 [1.56, 1.86] & --- & erasing & q-stat. \\
S6.4 (acq.~a) & 8192 & two-pause (0.45$\to$0.35) & 0.69 [0.69, 0.70] & 1.66 [1.56, 1.83] & --- & erasing & q-stat. \\
\midrule
S6.4 (acq.~b) & 14029 & single-pause (0.40, 100\,$\mu$s) & 2.19 [1.88, 2.87] & 1.43 [1.39, 1.47] & 30 [24, 39] & erasing & q-stat. \\
S6.4 (acq.~b) & 14029 & long-hold (0.35, 500\,$\mu$s) & 2.08 [1.91, 2.28] & --- & 37 [31, 39] & retaining & drift \\
S6.4 (acq.~b) & 14029 & two-pause (0.45$\to$0.35) & 2.20 [1.93, 2.67] & --- & 40 [24, 64] & retaining & q-stat. \\
S6.4 (acq.~b) & 8192 & single-pause (0.40, 100\,$\mu$s) & 0.63 [0.62, 0.64] & 1.95 [1.78, 2.27] & --- & erasing & q-stat. \\
S6.4 (acq.~b) & 8192 & long-hold (0.35, 500\,$\mu$s) & 1.33 [1.30, 1.37] & 2.10 [1.89, 12.00] & --- & erasing & q-stat. \\
S6.4 (acq.~b) & 8192 & two-pause (0.45$\to$0.35) & 0.69 [0.69, 0.70] & 2.01 [1.82, 12.00] & --- & erasing & q-stat. \\
\midrule
Advantage2 (native) & 20260421 & single-pause (0.40, 100\,$\mu$s) & --- & 6.92 [6.31, 8.14] & 131 [64, 556] & erasing & q-stat. \\
Advantage2 (native) & 20260421 & long-hold (0.35, 500\,$\mu$s) & --- & 7.07 [6.47, 8.14] & 131 [64, 556] & erasing & q-stat. \\
Advantage2 (native) & 20260421 & two-pause (0.45$\to$0.35) & --- & 6.78 [6.23, 7.74] & 131 [50, 3000] & erasing & q-stat. \\
Advantage2 (native) & 20260422 & single-pause (0.40, 100\,$\mu$s) & --- & 12.00 [12.00, 12.00] & 131 [50, 3000] & erasing & q-stat. \\
Advantage2 (native) & 20260422 & long-hold (0.35, 500\,$\mu$s) & --- & 4.00 [3.48, 12.00] & 131 [50, 3000] & erasing & q-stat. \\
Advantage2 (native) & 20260422 & two-pause (0.45$\to$0.35) & --- & 12.00 [12.00, 12.00] & 131 [50, 3000] & erasing & q-stat. \\
Advantage2 (native) & 20260423 & single-pause (0.40, 100\,$\mu$s) & --- & 2.16 [2.02, 2.32] & 0 [0, 3000] & erasing & q-stat. \\
Advantage2 (native) & 20260423 & long-hold (0.35, 500\,$\mu$s) & --- & --- & 131 [39, 556] & retaining & q-stat. \\
Advantage2 (native) & 20260423 & two-pause (0.45$\to$0.35) & --- & --- & 131 [15, 3000] & retaining & q-stat. \\
\bottomrule
\end{tabular}
\vspace{1ex}
\begin{minipage}{\linewidth}
\scriptsize
\textit{Notes.} All 39 Phase-3A cells are listed; no cell is omitted (\textsc{analysis\_spec} \S7.5). Devices: Advantage2\_system1 (\texttt{A2}), the two Advantage\_system6.4 acquisitions analysed separately (\texttt{S64a}, \texttt{S64b}), and the native-graph Advantage2 cells (\texttt{A2nat}). Intervals are 95\%: $\beta_\mathrm{mm}$ and $\hat\beta$ from the archived bootstrap, $n_\mathrm{eff}$ from grid inversion against the Clopper--Pearson interval of $S(49)$. $\beta_\mathrm{mm}$ is blank for the 9 native-graph cells, whose schedules have no Phase-2 counterpart (\S7.5). $\hat\beta$ is blank for the 12 memory-retaining cells, which carry no equilibrium fit (\S4.3, \S8.2). $n_\mathrm{eff}$ is blank for the 7 cells the artifact records as ``no match within scanned family''. This is an estimation-only table: no test, no multiplicity control (\S11). $n_\mathrm{eff}$ is an empirical clock calibration and licenses no mechanism claim (\S9.5, Table~\ref{tab:licensing}).
\end{minipage}
\end{table*}

%% file: tables/family_p2resp.tex
\begin{table*}[htbp]
\centering
\caption{Testing family P2-RESP (all cells). Per-instance schedule responsiveness, primary: single-pause (0.40, 100\,$\mu$s) vs.\ two-pause (0.45$\to$0.35). Estimate, 95\% CI, raw permutation $p$ and BH-adjusted $q$ for every one of the 20 instances (\textsc{analysis\_spec} \S6.1, \S11).}
\label{tab:family_p2resp}
\footnotesize
\begin{tabular}{lllllll}
\toprule
Instance & $T_\mathrm{obs}$ (TVD) & 95\% CI & raw $p$ & $q_\mathrm{BH}$ & $T^{95}_\mathrm{null}$ & Verdict \\
\midrule
42 & 0.6206 & [0.6029, 0.6384] & $1.0\times10^{-4}$ & $2.5\times10^{-4}$ & 0.1680 & responsive \\
123 & 0.0248 & [0.0072, 0.0491] & 0.0631 & 0.0971 & 0.0260 & quiet \\
456 & 0.0003 & [0.0001, 0.0005] & 0.0520 & 0.0867 & 0.0003 & quiet \\
789 & 0.0013 & [0.0002, 0.0029] & 0.1765 & 0.2353 & 0.0017 & quiet \\
1024 & 0.0657 & [0.0596, 0.0779] & $1.0\times10^{-4}$ & $2.5\times10^{-4}$ & 0.0313 & responsive \\
2048 & 0.0854 & [0.0061, 0.1755] & 0.0981 & 0.1401 & 0.1007 & quiet \\
4096 & 0.3774 & [0.3448, 0.4131] & $1.0\times10^{-4}$ & $2.5\times10^{-4}$ & 0.1135 & responsive \\
8192 & 0.2230 & [0.1901, 0.2528] & $1.0\times10^{-4}$ & $2.5\times10^{-4}$ & 0.0716 & responsive \\
10001 & 0.1286 & [0.0419, 0.2118] & 0.0058 & 0.0116 & 0.0942 & responsive \\
12011 & 0.1618 & [0.1304, 0.1930] & $1.0\times10^{-4}$ & $2.5\times10^{-4}$ & 0.0558 & responsive \\
14029 & 0.0030 & [0.0024, 0.1239] & 0.9664 & 0.9664 & 0.1081 & quiet \\
16033 & 0.0005 & [0.0002, 0.0009] & 0.0115 & 0.0209 & 0.0004 & responsive \\
18041 & 0.0265 & [0.0139, 0.1144] & 0.5677 & 0.5976 & 0.0882 & quiet \\
20051 & 0.0422 & [0.0330, 0.0512] & $1.0\times10^{-4}$ & $2.5\times10^{-4}$ & 0.0154 & responsive \\
22063 & 0.0006 & [0.0001, 0.0016] & 0.2956 & 0.3695 & 0.0010 & quiet \\
24071 & 0.0024 & [0.0016, 0.0034] & $1.0\times10^{-4}$ & $2.5\times10^{-4}$ & 0.0012 & responsive \\
26083 & 0.0077 & [0.0014, 0.0233] & 0.3575 & 0.4205 & 0.0151 & quiet \\
28087 & 0.0291 & [0.0193, 0.0456] & 0.0028 & 0.0062 & 0.0196 & responsive \\
30091 & 0.0196 & [0.0133, 0.0523] & 0.4625 & 0.5138 & 0.0414 & quiet \\
32099 & 0.0282 & [0.0228, 0.0363] & $1.0\times10^{-4}$ & $2.5\times10^{-4}$ & 0.0107 & responsive \\
\bottomrule
\end{tabular}
\vspace{1ex}
\begin{minipage}{\linewidth}
\scriptsize
\textit{Notes.} All 20 Phase-2 instances are reported; this family has no excluded cell. Instance 16033 is present here and absent from the Phase-3A per-cell families, where its Advantage2 mixing-time acquisition died mid-instance (\textsc{analysis\_spec} \S2.3). Estimate is $T_\mathrm{obs}$, the total variation distance between the gauge-restored pooled subsystem distributions of the two schedules; intervals are 95\% block bootstrap (B=10000) over 48 (gauge, bath, init) blocks. Raw $p$ is the block-label permutation test with $B = 10,000$ ($p = (1 + \#\{T_\mathrm{perm} \geq T_\mathrm{obs}\})/(B+1)$, so $p \geq 1/(B+1)$ by construction); $q_\mathrm{BH}$ is Benjamini--Hochberg within this 20-cell family at $\alpha = 0.05$. $T^{95}_\mathrm{null}$ is the 95th percentile of the permutation null. ``Responsive'' means BH-significant; the fixed $\mathrm{TVD} \geq 0.05$ threshold is retired (\S6.1). Both raw $p$ and $q$ are archived per cell; verdicts use $q$ (\S11). 11 of 20 cells are responsive.
\end{minipage}
\end{table*}

%% file: tables/family_p2resp_lh.tex
\begin{table*}[htbp]
\centering
\caption{Testing family P2-RESP-LH (all cells). Per-instance schedule responsiveness, long-hold contrast: single-pause (0.40, 100\,$\mu$s) vs.\ long-hold (0.35, 500\,$\mu$s). Estimate, 95\% CI, raw permutation $p$ and BH-adjusted $q$ for every one of the 20 instances (\textsc{analysis\_spec} \S6.2, \S11).}
\label{tab:family_p2resp_lh}
\footnotesize
\begin{tabular}{lllllll}
\toprule
Instance & $T_\mathrm{obs}$ (TVD) & 95\% CI & raw $p$ & $q_\mathrm{BH}$ & $T^{95}_\mathrm{null}$ & Verdict \\
\midrule
42 & 0.8720 & [0.8641, 0.8794] & $1.0\times10^{-4}$ & $1.8\times10^{-4}$ & 0.2235 & responsive \\
123 & 0.0449 & [0.0419, 0.0557] & $4.0\times10^{-4}$ & $6.7\times10^{-4}$ & 0.0272 & responsive \\
456 & 0.0002 & [0.0000, 0.0003] & 0.2356 & 0.2771 & 0.0002 & quiet \\
789 & 0.0020 & [0.0007, 0.0037] & 0.0424 & 0.0530 & 0.0019 & quiet \\
1024 & 0.2778 & [0.2523, 0.3032] & $1.0\times10^{-4}$ & $1.8\times10^{-4}$ & 0.0845 & responsive \\
2048 & 0.5642 & [0.4154, 0.6915] & $1.0\times10^{-4}$ & $1.8\times10^{-4}$ & 0.2105 & responsive \\
4096 & 0.3795 & [0.3469, 0.4148] & $1.0\times10^{-4}$ & $1.8\times10^{-4}$ & 0.1106 & responsive \\
8192 & 0.8370 & [0.8056, 0.8642] & $1.0\times10^{-4}$ & $1.8\times10^{-4}$ & 0.2316 & responsive \\
10001 & 0.5675 & [0.3984, 0.7214] & $1.0\times10^{-4}$ & $1.8\times10^{-4}$ & 0.2305 & responsive \\
12011 & 0.1603 & [0.1260, 0.1927] & $1.0\times10^{-4}$ & $1.8\times10^{-4}$ & 0.0572 & responsive \\
14029 & 0.0261 & [0.0023, 0.1215] & 0.6209 & 0.6536 & 0.0986 & quiet \\
16033 & 0.0005 & [0.0002, 0.0010] & 0.0204 & 0.0272 & 0.0004 & responsive \\
18041 & 0.1074 & [0.1004, 0.1504] & 0.0186 & 0.0266 & 0.0921 & responsive \\
20051 & 0.1067 & [0.0861, 0.1270] & $1.0\times10^{-4}$ & $1.8\times10^{-4}$ & 0.0360 & responsive \\
22063 & 0.0002 & [0.0001, 0.0010] & 0.8374 & 0.8374 & 0.0008 & quiet \\
24071 & 0.0022 & [0.0014, 0.0032] & $1.0\times10^{-4}$ & $1.8\times10^{-4}$ & 0.0011 & responsive \\
26083 & 0.0068 & [0.0017, 0.0219] & 0.4585 & 0.5094 & 0.0165 & quiet \\
28087 & 0.0276 & [0.0190, 0.0445] & 0.0032 & 0.0049 & 0.0191 & responsive \\
30091 & 0.1495 & [0.1228, 0.1778] & $1.0\times10^{-4}$ & $1.8\times10^{-4}$ & 0.0628 & responsive \\
32099 & 0.0302 & [0.0239, 0.0381] & $1.0\times10^{-4}$ & $1.8\times10^{-4}$ & 0.0110 & responsive \\
\bottomrule
\end{tabular}
\vspace{1ex}
\begin{minipage}{\linewidth}
\scriptsize
\textit{Notes.} All 20 Phase-2 instances are reported; this family has no excluded cell. Instance 16033 is present here and absent from the Phase-3A per-cell families, where its Advantage2 mixing-time acquisition died mid-instance (\textsc{analysis\_spec} \S2.3). Estimate is $T_\mathrm{obs}$, the total variation distance between the gauge-restored pooled subsystem distributions of the two schedules; intervals are 95\% block bootstrap (B=10000) over 48 (gauge, bath, init) blocks. Raw $p$ is the block-label permutation test with $B = 10,000$ ($p = (1 + \#\{T_\mathrm{perm} \geq T_\mathrm{obs}\})/(B+1)$, so $p \geq 1/(B+1)$ by construction); $q_\mathrm{BH}$ is Benjamini--Hochberg within this 20-cell family at $\alpha = 0.05$. $T^{95}_\mathrm{null}$ is the 95th percentile of the permutation null. ``Responsive'' means BH-significant; the fixed $\mathrm{TVD} \geq 0.05$ threshold is retired (\S6.1). Both raw $p$ and $q$ are archived per cell; verdicts use $q$ (\S11). 15 of 20 cells are responsive.
\end{minipage}
\end{table*}

%% file: tables/family_fact.tex
\begin{table*}[htbp]
\centering
\caption{Testing family FACT (all cells), depth effects. Depth$\times$duration factorial decomposition of $\beta_\mathrm{mm}$: estimate, 95\% CI, bootstrap two-sided $p$ and BH-adjusted $q$ for each of the 20 instances. The BH family is all 60 effects (\textsc{analysis\_spec} \S7.2, \S11).}
\label{tab:family_fact_depth}
\footnotesize
\begin{tabular}{lllllll}
\toprule
Instance & Effect on $\beta_\mathrm{mm}$ & 95\% CI & bootstrap $p$ & $q_\mathrm{BH}$ & CI excl.\ 0 & Verdict \\
\midrule
42 & 2.825 & [2.688, 3.001] & 0 & 0 & yes & significant \\
123 & -0.422 & [-0.505, -0.337] & 0 & 0 & yes & significant \\
456 & -0.233 & [-0.316, -0.149] & 0 & 0 & yes & significant \\
789 & -0.650 & [-1.090, -0.184] & 0.0090 & 0.0123 & yes & significant \\
1024 & 0.208 & [0.185, 0.232] & 0 & 0 & yes & significant \\
2048 & 0.150 & [0.105, 0.202] & 0 & 0 & yes & significant \\
4096 & 1.996 & [1.793, 2.226] & 0 & 0 & yes & significant \\
8192 & 0.400 & [0.382, 0.422] & 0 & 0 & yes & significant \\
10001 & 0.149 & [0.096, 0.224] & 0 & 0 & yes & significant \\
12011 & 2.737 & [2.595, 2.903] & 0 & 0 & yes & significant \\
14029 & -0.068 & [-0.535, 0.244] & 0.6380 & 0.6488 & no & not significant \\
16033 & 0.684 & [0.189, 1.327] & 0.0082 & 0.0115 & yes & significant \\
18041 & 0.200 & [0.170, 0.232] & 0 & 0 & yes & significant \\
20051 & -0.505 & [-0.606, -0.405] & 0 & 0 & yes & significant \\
22063 & -2.373 & [-2.555, -2.203] & 0 & 0 & yes & significant \\
24071 & -0.885 & [-1.187, -0.633] & 0 & 0 & yes & significant \\
26083 & -0.345 & [-0.592, -0.094] & 0.0076 & 0.0109 & yes & significant \\
28087 & 6.001 & [5.195, 6.789] & 0 & 0 & yes & significant \\
30091 & 0.276 & [0.185, 0.364] & 0 & 0 & yes & significant \\
32099 & 1.696 & [1.501, 1.897] & 0 & 0 & yes & significant \\
\bottomrule
\end{tabular}
\vspace{1ex}
\begin{minipage}{\linewidth}
\scriptsize
\textit{Notes.} Depth main effect: mean over $t_\mathrm{hold}$ of $[\beta_\mathrm{mm}(0.35,\cdot) - \beta_\mathrm{mm}(0.40,\cdot)]$, on the moment-matched inverse temperature $\beta_\mathrm{mm}$ of the Phase-2 $2\times2$ single-pause factorial $s_p \in \{0.35, 0.40\} \times t_\mathrm{hold} \in \{100, 500\}$. All 20 Phase-2 instances are reported, including 16033; this family has no excluded cell (instance 16033 is excluded only from the Phase-3A per-cell families, \textsc{analysis\_spec} \S2.3). Intervals are 95\% joint block bootstrap (B = 10{,}000) over 96 (gauge, bath, init) blocks. $q_\mathrm{BH}$ is Benjamini--Hochberg over the whole 60-effect family (all three effect types jointly, Tables~\ref{tab:family_fact_depth}--\ref{tab:family_fact_interaction}) at $\alpha = 0.05$; an effect is significant iff its 95\% CI excludes 0 after that adjustment (\S7.2). A $p$ or $q$ of exactly 0 is the archived value: no bootstrap replicate crossed zero (37 of 60 effects family-wide). 19 of 20 effects are significant here; 46 of 60 family-wide. Scope (mandatory, \S7.2): endpoint-only: decomposes endpoint response, not relaxation pathways; licenses no freeze-out mechanism claim (\S7.2).
\end{minipage}
\end{table*}

\begin{table*}[htbp]
\centering
\caption{Testing family FACT (all cells), duration effects. Depth$\times$duration factorial decomposition of $\beta_\mathrm{mm}$: estimate, 95\% CI, bootstrap two-sided $p$ and BH-adjusted $q$ for each of the 20 instances. The BH family is all 60 effects (\textsc{analysis\_spec} \S7.2, \S11).}
\label{tab:family_fact_duration}
\footnotesize
\begin{tabular}{lllllll}
\toprule
Instance & Effect on $\beta_\mathrm{mm}$ & 95\% CI & bootstrap $p$ & $q_\mathrm{BH}$ & CI excl.\ 0 & Verdict \\
\midrule
42 & 1.767 & [1.631, 1.940] & 0 & 0 & yes & significant \\
123 & -0.214 & [-0.299, -0.130] & 0 & 0 & yes & significant \\
456 & -0.271 & [-0.355, -0.190] & 0 & 0 & yes & significant \\
789 & -0.231 & [-0.684, 0.237] & 0.3272 & 0.3849 & no & not significant \\
1024 & 0.134 & [0.111, 0.158] & 0 & 0 & yes & significant \\
2048 & 0.110 & [0.065, 0.162] & 0 & 0 & yes & significant \\
4096 & 0.671 & [0.438, 0.876] & 0 & 0 & yes & significant \\
8192 & 0.298 & [0.279, 0.320] & 0 & 0 & yes & significant \\
10001 & 0.093 & [0.041, 0.169] & 0 & 0 & yes & significant \\
12011 & 1.035 & [0.888, 1.195] & 0 & 0 & yes & significant \\
14029 & -0.044 & [-0.467, 0.304] & 0.7804 & 0.7804 & no & not significant \\
16033 & 0.647 & [0.081, 1.201] & 0.0273 & 0.0363 & yes & significant \\
18041 & 0.158 & [0.127, 0.189] & 0 & 0 & yes & significant \\
20051 & -0.531 & [-0.633, -0.428] & 0 & 0 & yes & significant \\
22063 & -1.479 & [-1.660, -1.310] & 0 & 0 & yes & significant \\
24071 & -0.113 & [-0.393, 0.148] & 0.3974 & 0.4499 & no & not significant \\
26083 & 0.102 & [-0.148, 0.353] & 0.4164 & 0.4627 & no & not significant \\
28087 & 0.302 & [-0.512, 1.110] & 0.4408 & 0.4809 & no & not significant \\
30091 & 0.239 & [0.152, 0.329] & 0 & 0 & yes & significant \\
32099 & 0.590 & [0.391, 0.785] & 0 & 0 & yes & significant \\
\bottomrule
\end{tabular}
\vspace{1ex}
\begin{minipage}{\linewidth}
\scriptsize
\textit{Notes.} Duration main effect: mean over $s_p$ of $[\beta_\mathrm{mm}(\cdot,500) - \beta_\mathrm{mm}(\cdot,100)]$, on the moment-matched inverse temperature $\beta_\mathrm{mm}$ of the Phase-2 $2\times2$ single-pause factorial $s_p \in \{0.35, 0.40\} \times t_\mathrm{hold} \in \{100, 500\}$. All 20 Phase-2 instances are reported, including 16033; this family has no excluded cell (instance 16033 is excluded only from the Phase-3A per-cell families, \textsc{analysis\_spec} \S2.3). Intervals are 95\% joint block bootstrap (B = 10{,}000) over 96 (gauge, bath, init) blocks. $q_\mathrm{BH}$ is Benjamini--Hochberg over the whole 60-effect family (all three effect types jointly, Tables~\ref{tab:family_fact_depth}--\ref{tab:family_fact_interaction}) at $\alpha = 0.05$; an effect is significant iff its 95\% CI excludes 0 after that adjustment (\S7.2). A $p$ or $q$ of exactly 0 is the archived value: no bootstrap replicate crossed zero (37 of 60 effects family-wide). 15 of 20 effects are significant here; 46 of 60 family-wide. Scope (mandatory, \S7.2): endpoint-only: decomposes endpoint response, not relaxation pathways; licenses no freeze-out mechanism claim (\S7.2).
\end{minipage}
\end{table*}

\begin{table*}[htbp]
\centering
\caption{Testing family FACT (all cells), interaction effects. Depth$\times$duration factorial decomposition of $\beta_\mathrm{mm}$: estimate, 95\% CI, bootstrap two-sided $p$ and BH-adjusted $q$ for each of the 20 instances. The BH family is all 60 effects (\textsc{analysis\_spec} \S7.2, \S11).}
\label{tab:family_fact_interaction}
\footnotesize
\begin{tabular}{lllllll}
\toprule
Instance & Effect on $\beta_\mathrm{mm}$ & 95\% CI & bootstrap $p$ & $q_\mathrm{BH}$ & CI excl.\ 0 & Verdict \\
\midrule
42 & 3.218 & [2.945, 3.564] & 0 & 0 & yes & significant \\
123 & -0.224 & [-0.396, -0.060] & 0.0052 & 0.0078 & yes & significant \\
456 & 0.530 & [0.365, 0.695] & 0 & 0 & yes & significant \\
789 & -1.090 & [-1.963, -0.124] & 0.0292 & 0.0381 & yes & significant \\
1024 & 0.152 & [0.106, 0.201] & 0 & 0 & yes & significant \\
2048 & 0.147 & [0.057, 0.251] & $6.0\times10^{-4}$ & $9.2\times10^{-4}$ & yes & significant \\
4096 & -0.264 & [-0.760, 0.121] & 0.2029 & 0.2434 & no & not significant \\
8192 & 0.476 & [0.438, 0.520] & 0 & 0 & yes & significant \\
10001 & 0.140 & [0.033, 0.291] & 0.0074 & 0.0108 & yes & significant \\
12011 & -1.488 & [-1.786, -1.172] & 0 & 0 & yes & significant \\
14029 & -0.213 & [-0.983, 0.559] & 0.5308 & 0.5587 & no & not significant \\
16033 & -0.902 & [-2.087, 0.169] & 0.0954 & 0.1193 & no & not significant \\
18041 & 0.271 & [0.211, 0.334] & 0 & 0 & yes & significant \\
20051 & 0.549 & [0.349, 0.758] & 0 & 0 & yes & significant \\
22063 & -0.094 & [-0.438, 0.271] & 0.5976 & 0.6182 & no & not significant \\
24071 & 0.520 & [-0.018, 1.088] & 0.0578 & 0.0738 & no & not significant \\
26083 & -0.394 & [-0.889, 0.112] & 0.1232 & 0.1509 & no & not significant \\
28087 & -0.527 & [-2.164, 1.073] & 0.5306 & 0.5587 & no & not significant \\
30091 & 0.312 & [0.130, 0.494] & $6.0\times10^{-4}$ & $9.2\times10^{-4}$ & yes & significant \\
32099 & -0.181 & [-0.568, 0.213] & 0.3812 & 0.4398 & no & not significant \\
\bottomrule
\end{tabular}
\vspace{1ex}
\begin{minipage}{\linewidth}
\scriptsize
\textit{Notes.} Interaction: $\beta_\mathrm{mm}(0.35,500) - \beta_\mathrm{mm}(0.35,100) - \beta_\mathrm{mm}(0.40,500) + \beta_\mathrm{mm}(0.40,100)$, on the moment-matched inverse temperature $\beta_\mathrm{mm}$ of the Phase-2 $2\times2$ single-pause factorial $s_p \in \{0.35, 0.40\} \times t_\mathrm{hold} \in \{100, 500\}$. All 20 Phase-2 instances are reported, including 16033; this family has no excluded cell (instance 16033 is excluded only from the Phase-3A per-cell families, \textsc{analysis\_spec} \S2.3). Intervals are 95\% joint block bootstrap (B = 10{,}000) over 96 (gauge, bath, init) blocks. $q_\mathrm{BH}$ is Benjamini--Hochberg over the whole 60-effect family (all three effect types jointly, Tables~\ref{tab:family_fact_depth}--\ref{tab:family_fact_interaction}) at $\alpha = 0.05$; an effect is significant iff its 95\% CI excludes 0 after that adjustment (\S7.2). A $p$ or $q$ of exactly 0 is the archived value: no bootstrap replicate crossed zero (37 of 60 effects family-wide). 12 of 20 effects are significant here; 46 of 60 family-wide. Scope (mandatory, \S7.2): endpoint-only: decomposes endpoint response, not relaxation pathways; licenses no freeze-out mechanism claim (\S7.2).
\end{minipage}
\end{table*}

%% file: tables/family_eqres.tex
\begin{table*}[htbp]
\centering
\caption{Testing family EQ-RES (all cells). Per-cell equilibrium-family residual decision: best-fitting tested equilibrium family, its residual, materiality, raw $p$ and BH-adjusted $q$ for every one of the 27 memory-erasing cells (\textsc{analysis\_spec} \S8, \S11).}
\label{tab:family_eqres}
\tiny
\setlength{\tabcolsep}{3pt}
\begin{tabular}{llllllllll}
\toprule
Device & Instance & Schedule & $b^*$ & Best family & $\mathrm{TVD}_\mathrm{best}$ & Material & raw $p$ & $q_\mathrm{BH}$ & Verdict \\
\midrule
Advantage2 & 10001 & single-pause (0.40, 100\,$\mu$s) & 0 & classical Gibbs & 0.8052 & yes & $5.0\times10^{-4}$ & 0.0012 & resolved residual \\
Advantage2 & 10001 & long-hold (0.35, 500\,$\mu$s) & 0 & classical Gibbs & 0.8052 & yes & $5.0\times10^{-4}$ & 0.0012 & resolved residual \\
Advantage2 & 10001 & two-pause (0.45$\to$0.35) & 0 & classical Gibbs & 0.8052 & yes & $5.0\times10^{-4}$ & 0.0012 & resolved residual \\
Advantage2 & 12011 & two-pause (0.45$\to$0.35) & 16 & TFIM Gibbs & 0.0391 & no & $5.0\times10^{-4}$ & 0.0012 & none \\
Advantage2 & 14029 & single-pause (0.40, 100\,$\mu$s) & 0 & classical Gibbs & 0.0010 & no & 0.6587 & 0.8084 & none \\
Advantage2 & 14029 & two-pause (0.45$\to$0.35) & 0 & classical Gibbs & 0.0377 & no & 0.0100 & 0.0208 & none \\
Advantage2 & 2048 & two-pause (0.45$\to$0.35) & 6 & classical Gibbs & 0.7707 & yes & $5.0\times10^{-4}$ & 0.0012 & resolved residual \\
Advantage2 & 4096 & single-pause (0.40, 100\,$\mu$s) & 0 & classical Gibbs & 0.0159 & no & $5.0\times10^{-4}$ & 0.0012 & none \\
Advantage2 & 4096 & long-hold (0.35, 500\,$\mu$s) & 0 & classical Gibbs & $4.8\times10^{-16}$ & no & 1.0000 & 1.0000 & none \\
Advantage2 & 4096 & two-pause (0.45$\to$0.35) & 0 & classical Gibbs & $4.8\times10^{-16}$ & no & 1.0000 & 1.0000 & none \\
Advantage2 & 8192 & single-pause (0.40, 100\,$\mu$s) & 0 & TFIM Gibbs & 0.0958 & yes & $5.0\times10^{-4}$ & 0.0012 & resolved residual \\
Advantage2 & 8192 & long-hold (0.35, 500\,$\mu$s) & 0 & classical Gibbs & $2.2\times10^{-4}$ & no & 0.0719 & 0.1078 & none \\
Advantage2 & 8192 & two-pause (0.45$\to$0.35) & 0 & TFIM Gibbs & 0.0176 & no & $5.0\times10^{-4}$ & 0.0012 & none \\
\midrule
S6.4 (acq.~a) & 8192 & single-pause (0.40, 100\,$\mu$s) & 0 & TFIM Gibbs & 0.0017 & no & 0.0150 & 0.0270 & none \\
S6.4 (acq.~a) & 8192 & long-hold (0.35, 500\,$\mu$s) & 0 & TFIM Gibbs & 0.0013 & no & 0.0111 & 0.0214 & none \\
S6.4 (acq.~a) & 8192 & two-pause (0.45$\to$0.35) & 0 & classical Gibbs & 0.0012 & no & 0.0319 & 0.0538 & none \\
\midrule
S6.4 (acq.~b) & 14029 & single-pause (0.40, 100\,$\mu$s) & 0 & TFIM Gibbs & 0.1248 & yes & $5.0\times10^{-4}$ & 0.0012 & resolved residual \\
S6.4 (acq.~b) & 8192 & single-pause (0.40, 100\,$\mu$s) & 0 & TFIM Gibbs & $7.4\times10^{-4}$ & no & 0.0041 & 0.0092 & none \\
S6.4 (acq.~b) & 8192 & long-hold (0.35, 500\,$\mu$s) & 0 & classical Gibbs & $2.6\times10^{-4}$ & no & 0.1729 & 0.2334 & none \\
S6.4 (acq.~b) & 8192 & two-pause (0.45$\to$0.35) & 0 & TFIM Gibbs & $8.6\times10^{-5}$ & no & 0.8936 & 1.0000 & none \\
\midrule
Advantage2 (native) & 20260421 & single-pause (0.40, 100\,$\mu$s) & 0 & TFIM Gibbs & 0.0012 & no & $5.0\times10^{-4}$ & 0.0012 & none \\
Advantage2 (native) & 20260421 & long-hold (0.35, 500\,$\mu$s) & 0 & TFIM Gibbs & $4.3\times10^{-4}$ & no & 0.1054 & 0.1498 & none \\
Advantage2 (native) & 20260421 & two-pause (0.45$\to$0.35) & 0 & TFIM Gibbs & $5.8\times10^{-4}$ & no & 0.0478 & 0.0759 & none \\
Advantage2 (native) & 20260422 & single-pause (0.40, 100\,$\mu$s) & 0 & classical Gibbs & $1.1\times10^{-11}$ & no & 1.0000 & 1.0000 & none \\
Advantage2 (native) & 20260422 & long-hold (0.35, 500\,$\mu$s) & 0 & classical Gibbs & $1.3\times10^{-6}$ & no & 0.5822 & 0.7486 & none \\
Advantage2 (native) & 20260422 & two-pause (0.45$\to$0.35) & 0 & classical Gibbs & $1.1\times10^{-11}$ & no & 1.0000 & 1.0000 & none \\
Advantage2 (native) & 20260423 & single-pause (0.40, 100\,$\mu$s) & 1 & TFIM Gibbs & 0.1279 & yes & $5.0\times10^{-4}$ & 0.0012 & resolved residual \\
\bottomrule
\end{tabular}
\vspace{1ex}
\begin{minipage}{\linewidth}
\scriptsize
\textit{Notes.} Realized family size 27 cells, recorded in \texttt{burnin/burnin\_per\_cell.json} before any fit was run (\textsc{analysis\_spec} \S8.2). Excluded: 12 memory-retaining cells, reason ``non-equilibrium by memory criterion'' --- they are non-equilibrium by the \S4.3 memory criterion and would trivially inflate the count. Excluded cells: A2/12011/single-pause; A2/12011/long-hold; A2/14029/long-hold; A2/2048/single-pause; A2/2048/long-hold; S6.4a/14029/single-pause; S6.4a/14029/long-hold; S6.4a/14029/two-pause; S6.4b/14029/long-hold; S6.4b/14029/two-pause; A2 (nat.)/20260423/long-hold; A2 (nat.)/20260423/two-pause. Instance 16033 and the aborted native stub run \texttt{phase3a\_native\_graph\_20260421\_211931} never enter this family (\S2.3); the retired SQA bracket is provenance-only and is not a fit family here (\S2.4, \S8.1). $b^*$ is the burn-in cycle (\S4.3). $\mathrm{TVD}_\mathrm{best}$ is the residual of the best-fitting tested equilibrium family; ``material'' is that residual against the 0.05 threshold (\S12 reports $\{0.03, 0.05, 0.075\}$). Raw $p$ is the per-cell fit test and $q_\mathrm{BH}$ is Benjamini--Hochberg within this family at $\alpha = 0.05$; a resolved residual requires both BH significance and materiality. 7 of 27 cells carry a resolved residual. Family verdict (\S8.5): RESOLVED BUT UNREPLICATED: resolved residuals exist but none satisfies the \S8.5 replication rule; reported as secondary, may not enter the abstract.
\end{minipage}
\end{table*}

%% file: tables/family_dynclass.tex
\begin{table*}[htbp]
\centering
\caption{Testing family DYN-CLASS (all cells), part 1 of 2: Advantage2 embedded and native-graph cells. Per-cell QPU survival $S(49)$ with 95\% CI, tested classical arm count, best-matching arm, intersection--union $p$, BH-adjusted $q$ and verdict for 27 of the 33 cells (\textsc{analysis\_spec} \S9.4, \S11).}
\label{tab:family_dynclass}
\tiny
\setlength{\tabcolsep}{2.5pt}
\begin{tabular}{lllllllllll}
\toprule
Device & Instance & Schedule & QPU $S(49)$ [95\% CI] & Basis & Arms & Best-matching arm & Arm test & $p_\mathrm{cell}$ & $q_\mathrm{BH}$ & Verdict \\
\midrule
A2 & 10001 & single-pause & 1.000 [0.960, 1.000] & $n=90$ & 489 & \parbox[t]{0.145\textwidth}{\raggedright metropolis(\allowbreak{}beta=3, a=707.2)} & Fisher exact & 1.0000 & 1.0000 & reproduced \\
A2 & 10001 & long-hold & 1.000 [0.960, 1.000] & $n=90$ & 489 & \parbox[t]{0.145\textwidth}{\raggedright metropolis(\allowbreak{}beta=3, a=707.2)} & Fisher exact & 1.0000 & 1.0000 & reproduced \\
A2 & 10001 & two-pause & 1.000 [0.960, 1.000] & $n=90$ & 489 & \parbox[t]{0.145\textwidth}{\raggedright metropolis(\allowbreak{}beta=3, a=707.2)} & Fisher exact & 1.0000 & 1.0000 & reproduced \\
A2 & 12011 & single-pause & 0.544 [0.436, 0.650] & $n=90$ & 489 & \parbox[t]{0.145\textwidth}{\raggedright metropolis(\allowbreak{}beta=3, a=1145)} & Fisher exact & 1.0000 & 1.0000 & reproduced \\
A2 & 12011 & long-hold & 0.556 [0.447, 0.660] & $n=90$ & 489 & \parbox[t]{0.145\textwidth}{\raggedright metropolis(\allowbreak{}beta=3, a=1145)} & Fisher exact & 1.0000 & 1.0000 & reproduced \\
A2 & 12011 & two-pause & 0.556 [0.447, 0.660] & $n=90$ & 489 & \parbox[t]{0.145\textwidth}{\raggedright metropolis(\allowbreak{}beta=3, a=1145)} & Fisher exact & 1.0000 & 1.0000 & reproduced \\
A2 & 14029 & single-pause & 0.989 [0.940, 1.000] & $n=90$ & 583 & \parbox[t]{0.145\textwidth}{\raggedright metropolis(\allowbreak{}beta=3.55, a=166.7)} & Fisher exact & 1.0000 & 1.0000 & reproduced \\
A2 & 14029 & long-hold & 0.835 [0.771, 0.884] & DL $n=450^\dagger$ & 583 & \parbox[t]{0.145\textwidth}{\raggedright SVMC(\allowbreak{}T=0.15GHz, clock=0.01sw/us)} & DL logit $z$ & 0.9680 & 1.0000 & reproduced \\
A2 & 14029 & two-pause & 0.864 [0.800, 0.910] & DL $n=450^\dagger$ & 583 & \parbox[t]{0.145\textwidth}{\raggedright metropolis(\allowbreak{}beta=2, a=103)} & DL logit $z$ & 0.7614 & 1.0000 & reproduced \\
A2 & 2048 & single-pause & 0.989 [0.940, 1.000] & $n=90$ & 381 & \parbox[t]{0.145\textwidth}{\raggedright metropolis(\allowbreak{}beta=4.29, a=899.8)} & Fisher exact & 1.0000 & 1.0000 & reproduced \\
A2 & 2048 & long-hold & 0.833 [0.740, 0.904] & $n=90$ & 381 & \parbox[t]{0.145\textwidth}{\raggedright metropolis(\allowbreak{}beta=7.219, a=707.2)} & Fisher exact & 1.0000 & 1.0000 & reproduced \\
A2 & 2048 & two-pause & 0.989 [0.940, 1.000] & $n=90$ & 381 & \parbox[t]{0.145\textwidth}{\raggedright metropolis(\allowbreak{}beta=4.29, a=899.8)} & Fisher exact & 1.0000 & 1.0000 & reproduced \\
A2 & 4096 & single-pause & 0.715 [0.671, 0.755] & DL $n=450^\dagger$ & 536 & \parbox[t]{0.145\textwidth}{\raggedright metropolis(\allowbreak{}beta=3, a=131)} & DL logit $z$ & 0.9161 & 1.0000 & reproduced \\
A2 & 4096 & long-hold & 1.000 [0.960, 1.000] & $n=90$ & 536 & \parbox[t]{0.145\textwidth}{\raggedright metropolis(\allowbreak{}beta=12, a=103)} & Fisher exact & 1.0000 & 1.0000 & reproduced \\
A2 & 4096 & two-pause & 1.000 [0.960, 1.000] & $n=90$ & 536 & \parbox[t]{0.145\textwidth}{\raggedright metropolis(\allowbreak{}beta=12, a=103)} & Fisher exact & 1.0000 & 1.0000 & reproduced \\
A2 & 8192 & single-pause & 0.288 [0.215, 0.375] & DL $n=450^\dagger$ & 778 & \parbox[t]{0.145\textwidth}{\raggedright metropolis(\allowbreak{}beta=1.25, a=19.08)} & DL logit $z$ & 0.9735 & 1.0000 & reproduced \\
A2 & 8192 & long-hold & 1.000 [0.960, 1.000] & $n=90$ & 778 & \parbox[t]{0.145\textwidth}{\raggedright PT(\allowbreak{}L\_4p29, 200 sweeps/cycle)} & Fisher exact & 1.0000 & 1.0000 & reproduced \\
A2 & 8192 & two-pause & 0.578 [0.469, 0.681] & $n=90$ & 778 & \parbox[t]{0.145\textwidth}{\raggedright metropolis(\allowbreak{}beta=0.9, a=436.9)} & Fisher exact & 1.0000 & 1.0000 & reproduced \\
\midrule
A2 (nat.) & 20260421 & single-pause & 1.000 [0.960, 1.000] & $n=90$ & 583 & \parbox[t]{0.145\textwidth}{\raggedright PT(\allowbreak{}L\_7p219, 5 sweeps/cycle)} & Fisher exact & 1.0000 & 1.0000 & reproduced \\
A2 (nat.) & 20260421 & long-hold & 1.000 [0.960, 1.000] & $n=90$ & 583 & \parbox[t]{0.145\textwidth}{\raggedright PT(\allowbreak{}L\_7p219, 5 sweeps/cycle)} & Fisher exact & 1.0000 & 1.0000 & reproduced \\
A2 (nat.) & 20260421 & two-pause & 0.989 [0.940, 1.000] & $n=90$ & 583 & \parbox[t]{0.145\textwidth}{\raggedright metropolis(\allowbreak{}beta=7.219, a=131)} & Fisher exact & 1.0000 & 1.0000 & reproduced \\
A2 (nat.) & 20260422 & single-pause & 1.000 [0.960, 1.000] & $n=90$ & 428 & \parbox[t]{0.145\textwidth}{\raggedright metropolis(\allowbreak{}beta=12, a=63.62)} & Fisher exact & 1.0000 & 1.0000 & reproduced \\
A2 (nat.) & 20260422 & long-hold & 1.000 [0.960, 1.000] & $n=90$ & 428 & \parbox[t]{0.145\textwidth}{\raggedright metropolis(\allowbreak{}beta=12, a=63.62)} & Fisher exact & 1.0000 & 1.0000 & reproduced \\
A2 (nat.) & 20260422 & two-pause & 1.000 [0.960, 1.000] & $n=90$ & 428 & \parbox[t]{0.145\textwidth}{\raggedright metropolis(\allowbreak{}beta=12, a=63.62)} & Fisher exact & 1.0000 & 1.0000 & reproduced \\
A2 (nat.) & 20260423 & single-pause & 0.100 [0.047, 0.181] & $n=90$ & 489 & \parbox[t]{0.145\textwidth}{\raggedright metropolis(\allowbreak{}beta=0.25, a=269.9)} & Fisher exact & 1.0000 & 1.0000 & reproduced \\
A2 (nat.) & 20260423 & long-hold & 0.444 [0.340, 0.553] & $n=90$ & 489 & \parbox[t]{0.145\textwidth}{\raggedright PT(\allowbreak{}L\_betahat\_2p15, 50 sweeps/cycle)} & Fisher exact & 1.0000 & 1.0000 & reproduced \\
A2 (nat.) & 20260423 & two-pause & 0.378 [0.278, 0.486] & $n=90$ & 489 & \parbox[t]{0.145\textwidth}{\raggedright metropolis(\allowbreak{}beta=7.219, a=131)} & Fisher exact & 1.0000 & 1.0000 & reproduced \\
\bottomrule
\end{tabular}
\vspace{1ex}
\begin{minipage}{\linewidth}
\tiny
\textit{Notes.} Schedules: single-pause $(s_p = 0.40, 100\,\mu\mathrm{s})$, long-hold $(0.35, 500\,\mu\mathrm{s})$, two-pause $(0.45 \to 0.35)$. ``Arms'' counts the tested classical arms (Metropolis budget scan, regenerated PT, schedule-matched SVMC); ``best-matching arm'' is the archived \texttt{closest\_matching\_arm}, the arm attaining the cell $p$. $p_\mathrm{cell} = \max$ over arms (intersection--union: the null ``at least one tested arm reproduces the QPU'' is rejected only if every arm differs, \S9.4); $q_\mathrm{BH}$ is Benjamini--Hochberg over the 33-cell family at $\alpha = 0.05$ and verdicts use $q$ (\S11). ``Reproduced'' abbreviates ``not reproduced within the tested classical family''; 0 of 33 cells reject. The forbidden reading is ``classical dynamics excluded / quantum advantage (\S9.5)'' --- see Table~\ref{tab:licensing}. Excluded from this family (\S2.3--2.4): instance 16033, whose Advantage2 acquisition died mid-instance (90/76/0 chains); the aborted native stub \texttt{phase3a\_native\_graph\_20260421\_211931}, superseded by \texttt{...\_212042}; and the retired SQA arms, provenance-only after the $\Gamma$-units defect. $^\dagger$ For the 4 embedding-sweep-covered cells the DerSimonian--Laird estimate pooled over 5 placements ($n = 450$, from \texttt{embedding/random\_effects.json}) is \emph{the} estimate for every purpose and their arms are compared by the DL logit $z$-test, not Fisher exact; the single-placement $n = 90$ values appear only as forest-plot points (\S3.2 supersession, \S9.4). The six Advantage\_system6.4 cells are in Table~\ref{tab:family_dynclass_s64}.
\end{minipage}
\end{table*}

\begin{table*}[htbp]
\centering
\caption{Testing family DYN-CLASS (all cells), part 2 of 2: the 6 Advantage\_system6.4 cells, one row per acquisition. Cell membership in the family is per cell, not per acquisition (\textsc{analysis\_spec} \S3.2, \S9.4, \S11).}
\label{tab:family_dynclass_s64}
\tiny
\setlength{\tabcolsep}{2.5pt}
\begin{tabular}{lllllllllll}
\toprule
Instance & Schedule & Acq. & $S(49)$ [95\% CI] & Arms & Best-matching arm & Arm test & acq.\ $p$ & rollup $p_\mathrm{cell}$ & $q_\mathrm{BH}$ & Verdict \\
\midrule
14029 & single-pause & a & 0.522 [0.414, 0.629] & 583 & \parbox[t]{0.145\textwidth}{\raggedright metropolis(\allowbreak{}beta=3.55, a=19.08)} & Fisher exact & 1.0000 & 1.0000 & 1.0000 & reproduced \\
14029 & single-pause & b & 0.622 [0.514, 0.722] & 583 & \parbox[t]{0.145\textwidth}{\raggedright SVMC(\allowbreak{}T=0.45GHz, clock=0.03162sw/us)} & Fisher exact & 1.0000 & 1.0000 & 1.0000 & reproduced \\
14029 & long-hold & a & 0.611 [0.503, 0.712] & 583 & \parbox[t]{0.145\textwidth}{\raggedright metropolis(\allowbreak{}beta=1.5, a=39.3)} & Fisher exact & 1.0000 & 1.0000 & 1.0000 & reproduced \\
14029 & long-hold & b & 0.644 [0.537, 0.743] & 583 & \parbox[t]{0.145\textwidth}{\raggedright metropolis(\allowbreak{}beta=3, a=30.89)} & Fisher exact & 0.9071 & 1.0000 & 1.0000 & reproduced \\
14029 & two-pause & a & 0.589 [0.480, 0.692] & 583 & \parbox[t]{0.145\textwidth}{\raggedright PT(\allowbreak{}L\_4p29, 1 sweeps/cycle)} & Fisher exact & 1.0000 & 1.0000 & 1.0000 & reproduced \\
14029 & two-pause & b & 0.589 [0.480, 0.692] & 583 & \parbox[t]{0.145\textwidth}{\raggedright PT(\allowbreak{}L\_4p29, 1 sweeps/cycle)} & Fisher exact & 1.0000 & 1.0000 & 1.0000 & reproduced \\
8192 & single-pause & a & 0.989 [0.940, 1.000] & 778 & \parbox[t]{0.145\textwidth}{\raggedright PT(\allowbreak{}L\_4p29, 200 sweeps/cycle)} & Fisher exact & 1.0000 & 1.0000 & 1.0000 & reproduced \\
8192 & single-pause & b & 1.000 [0.960, 1.000] & 778 & \parbox[t]{0.145\textwidth}{\raggedright PT(\allowbreak{}L\_4p29, 200 sweeps/cycle)} & Fisher exact & 1.0000 & 1.0000 & 1.0000 & reproduced \\
8192 & long-hold & a & 0.989 [0.940, 1.000] & 778 & \parbox[t]{0.145\textwidth}{\raggedright PT(\allowbreak{}L\_4p29, 200 sweeps/cycle)} & Fisher exact & 1.0000 & 1.0000 & 1.0000 & reproduced \\
8192 & long-hold & b & 1.000 [0.960, 1.000] & 778 & \parbox[t]{0.145\textwidth}{\raggedright PT(\allowbreak{}L\_4p29, 200 sweeps/cycle)} & Fisher exact & 1.0000 & 1.0000 & 1.0000 & reproduced \\
8192 & two-pause & a & 0.989 [0.940, 1.000] & 778 & \parbox[t]{0.145\textwidth}{\raggedright PT(\allowbreak{}L\_4p29, 200 sweeps/cycle)} & Fisher exact & 1.0000 & 1.0000 & 1.0000 & reproduced \\
8192 & two-pause & b & 0.989 [0.940, 1.000] & 778 & \parbox[t]{0.145\textwidth}{\raggedright PT(\allowbreak{}L\_4p29, 200 sweeps/cycle)} & Fisher exact & 1.0000 & 1.0000 & 1.0000 & reproduced \\
\bottomrule
\end{tabular}
\vspace{1ex}
\begin{minipage}{\linewidth}
\tiny
\textit{Notes.} Schedules: single-pause $(s_p = 0.40, 100\,\mu\mathrm{s})$, long-hold $(0.35, 500\,\mu\mathrm{s})$, two-pause $(0.45 \to 0.35)$. ``Arms'' counts the tested classical arms (Metropolis budget scan, regenerated PT, schedule-matched SVMC); ``best-matching arm'' is the archived \texttt{closest\_matching\_arm}, the arm attaining the cell $p$. $p_\mathrm{cell} = \max$ over arms (intersection--union: the null ``at least one tested arm reproduces the QPU'' is rejected only if every arm differs, \S9.4); $q_\mathrm{BH}$ is Benjamini--Hochberg over the 33-cell family at $\alpha = 0.05$ and verdicts use $q$ (\S11). ``Reproduced'' abbreviates ``not reproduced within the tested classical family''; 0 of 33 cells reject. The forbidden reading is ``classical dynamics excluded / quantum advantage (\S9.5)'' --- see Table~\ref{tab:licensing}. Excluded from this family (\S2.3--2.4): instance 16033, whose Advantage2 acquisition died mid-instance (90/76/0 chains); the aborted native stub \texttt{phase3a\_native\_graph\_20260421\_211931}, superseded by \texttt{...\_212042}; and the retired SQA arms, provenance-only after the $\Gamma$-units defect. Each Advantage\_system6.4 cell is one member of the 33-cell family but carries two acquisitions, analysed separately at $n = 90$ each (\S3.2); both rows of a pair share the cell-level rollup $p$, $q$ and verdict. Rollup rule (\S3.2): requires both S6.4 acquisitions to reject (max of the two IUT p). The remaining 27 cells are in Table~\ref{tab:family_dynclass}.
\end{minipage}
\end{table*}